\documentclass{JHEP3}
\usepackage{amssymb,amsmath}
\usepackage{slashed,bm,bbm}
\usepackage{url}
\usepackage{graphicx,psfrag}
\newcommand{\half}{\frac{1}{2}}

\def\veps{\varepsilon}
\def\vp{\varphi}
\DeclareMathOperator{\tr}{tr}
\newcommand{\erw}[1]{\langle #1\rangle}
\def\vepsb{\bar{\varepsilon}}
\def\Son{S_{\rm on}}
\def\psib{\bar{\psi}}
\def\pa{\partial}
\def\a{\alpha}
\def\t{\theta}
\def\tb{\bar{\theta}}
\def\Qb{\bar{Q}}
\def\Db{\bar{D}}

\def\p{\phi}
\def\P{\Phi}

\def\d{\delta}
\def\psibar{\bar{\psi}}
\def\psb{\bar{\psi}}

\def\3{ss}

\def\Db{\bar{D}}
\def\inte{\hspace{-0.2cm}\int\hspace{-0.2cm}}
\def\dttb{d\t\hspace{-0.02cm}d\tb\hspace{0.02cm}}
\newcommand{\one}[1]{#1^{(1)}}
\newcommand{\two}[1]{#1^{(2)}}
\newcommand{\three}[1]{#1^{(3)}}
\newcommand{\nth}[1]{#1^{(n_\text{f})}}

\def\leftder{\nabla^{(+)}}
\def\rightder{\nabla^{(-)}}
\def\symder{\nabla^{(s)}}
\def\derSLAC{\nabla^{\text{SLAC}}}

\def\MW{m^{(W)}}

\def\n{\nabla}
\def\eq#1{(\ref{#1})}

\def\mb{m_b}
\def\mf{m_f}

\def\D{\mathrm{D}}
\newcommand{\nf}{n_\text{f}}
\newcommand{\Ct}{\tilde{C}}
\newtheorem{lemma}{Lemma} 
\def\c{\chi}
\def\Fc{{\cal F}^{(\nf)}}
\def\Ft{{\cal F}^{(4)}}
\newlength{\grwidth}
\providecommand{\href}[2]{#2}
\newcommand{\arxiv}[1]{[arXiv:\href{http://arxiv.org/abs/#1}{{\tt #1}}]}
\newcommand{\urlref}[1]{\href{#1}{\url{#1}}}
\title{Complete supersymmetry on the lattice and a No-Go theorem} 
\author{Georg Bergner\\
Theoretisch-Physikalisches Institut, Friedrich-Schiller-Universit{\"a}t Jena,
Max-Wien-Platz 1, D-07743 Jena, Germany\\
   E-mail: \email{g.bergner@tpi.uni-jena.de}}

\abstract{
In this work a lattice formulation of a supersymmetric theory is proposed and tested that preserves the complete supersymmetry on the lattice.
The results of a onedimensional nonperturbative simulation show the realization of the full supersymmetry and the correct continuum limit of the theory.
It is proven here that the violation of supersymmetry due to the absence of the Leibniz rule on the lattice can be amended only with a nonlocal derivative and nonlocal interaction term.
The fermion doubling problem is also discussed, which leads to another important source of supersymmetry breaking on the lattice.
This problem is also solved with a nonlocal realization.}
  
 \preprint{}
  
\begin{document}
\section{Introduction}
Supersymmetric theories have drawn much attention over the last decades.
There is a variety of application for this symmetry ranging from supersymmetric quantum mechanical systems to extensions of the standard model (like MSSM) and string theory.
While perturbative calculations and some nonperturbative arguments already relveal some intresting features of supersymmetric theories, it is still desirable to have a nonperturbative tool at hand for further investigations.
Lattice calculations have proven to provide an important insight in the nonperturbative sector of quantum field theories like QCD.
To get reliable results of supersymmetric models with this method the theory must be discretised in such a way that the correct theory is reproduced in the continuum limit.
The established techniques of discretisation lead, however, to a breaking of supersymmetry on the lattice.
There are various attempts to get, nevertheless, a supersymmetric theory in the continuum limit (cf.\ \cite{Giedt:2006pd,Feo:2002yi, Kaplan:2007zz} for reviews).
These approaches contain a fine-tuning towards the correct continuum limit \cite{Montvay:1995rs} and a partial realization of supersymmetry on the lattice \cite{Elitzur1982165,Catterall:2001wx,Kaplan:2002wv}.
However, the violation of supersymmetry at a finite lattice spacing seems to be unavoidable.
The reasons for this violation of the continuum supersymmetry on the lattice are further investigated in this work.
As a possible solution nonlocal realizations will be investigated.

The most difficult problem is the absence of the Leibniz rule that is necessary for the invariance of the (interacting) lattice action.
This problem was first discussed in \cite{Dondi:1976tx}, where as a possible solution an action with a nonlocal interaction term was presented.
This approach was further pursued in  \cite{Nojiri:1984ys,Nojiri:1985vb}, where also the breaking of the translational invariance of this suggestion was found.
In \cite{Bartels:1982ue} a discretisation of the four-dimensional Wess-Zumino model was suggested that respects all the continuum supersymmetry.
Its disadvantage is a nonlocal derivative operator in addition to the nonlocal interaction term in the lattice action. 
In \cite{Kato:2008fr} a ``No-Go theorem'' was presented, that for the implementation of full supersymmetry on the lattice either a nonlocal interaction term or a nonlocal derivative is necessary.
According to this argument the nonlocal SLAC derivative should provide an realization with intact supersymmetry on the lattice.
Such an action was already investigated in \cite{Bergner:2007pu}, where despite the usage of the nonlocal derivative operator still a violation of supersymmetry at a finite lattice spacing was found.
I will present here a more complete version of the simple ``No-Go'' statement. I will show that only with a nonlocal derivative \emph{and} a nonlocal interaction term supersymmetry can be realized on the lattice.
The conditions for a controlled approximation of the continuum theory on the lattice, presented here, also exclude the suggestion of \cite{Dondi:1976tx}.

For the evaluation of the importance of the supersymmetry breaking by all local lattice realization, I will compare this effect with other sources of supersymmetry breaking on the lattice.
These other important sources of supersymmetry breaking can be traced back to the fermion doubling problem.
Although the considerations are quite generic for supersymmetric lattice realizations, I will focus here on Wess-Zumino type models.

According to the ``No-Go'' statement the two options for the lattice simulations are that either locality or supersymmetry is not realized at a finite lattice spacing and must be recovered in the continuum limit.
The best way would be to compare the results of  both of these options. 
If they agree a supersymmetric and local continuum limit can be ensured in the nonperturbative sector theory.

It demands, however, a large numerical effort to simulate the nonlocal supersymmetric lattice actions.
I will present here a lattice formulation that allows for a simulation of at least the lowerdimensional models.
The first results of a such a fully supersymmetric lattice simulation are shown in this paper.

The paper is organized as follows: First I will give a short summary of supersymmetric models in the continuum, that are the starting point of the general discussion. 
In section \ref{secFailLeibn} I will introduce a lattice discretisation for supersymmetric theories and show how a violation of supersymmetry appears due to the absence of the Leibniz rule on the lattice. 
In section \ref{sec:NoGo} a simple ``No-Go'' statement for local supersymmetric lattice realizations will be presented.
A concrete lattice realization that allows for a simulation with intact supersymmetry on the lattice will be considered in section \ref{sec:Real}.
In most of the earlier work the supersymmetry breaking due to either the Leibniz rule or the fermion doubling problem was discussed.
For a more complete discussion and a comparison of these effects, section \ref{sec:doubling} adds some considerations of the doubling problem.
Section \ref{sec:PertTh} contains some perturbative arguments for the local continuum limit of the nonlocal lattice actions. The more important nonperturbative results of a supersymmetric lattice simulation are presented in section \ref{sec:Simulations}. The conclusions \ref{sec:Conclu} at the end of the paper contain some suggestions for further improvements of the full supersymmetric lattice simulations.
This work is based on an updated version of the results in \cite{Bergner_PhD_2009}.

\subsection{Continuum supersymmetry}
The results of this work are quite generic for supersymmetric field theories, i.~e.\ field theories with a symmetry that connects fermions and bosons and contain derivative operators in the symmetry transformations. 
Gauge theories demand some further discussions, especially when nonlocal operators are considered.
Therefore, the main focus of the paper are Wess-Zumino type models.
Models of that type are used for the matter sector of supersymmetric extensions of the standard model. 
The simplest example is the one-dimensional Wess-Zumino model. 
A derivation of the action in superspace and its off-shell representation is shown in appendix \ref{app:superspace}.
The on-shell action of this model is
\begin{equation}
\label{eq:cont1D}
 S=\inte dt \left( \half (\pa_t\vp)^2+\psib\slashed{\pa} \psi+\half (W'(\vp))^2+\psib W''(\vp)\psi\right)\,.
\end{equation}
It contains the scalar bosonic field $\vp$ and a complex fermion field $\psi$.
The bosonic potential, the Yukawa interaction between fermions and bosons, and possible mass terms are derived from the superpotential $W(\vp)$ that is a polynomial of the fields.\footnote{The prime denotes a derivative of the superpotential ($W'(\vp(t))=\frac{d}{d\vp(t)} W(\vp(t))$).}
Hence, there are relations between the bosonic and fermionic vertices.
A possible mass parameter appears in certain bosonic interaction vertices.
The action transforms under the supersymmetry transformations
\begin{equation}
\label{eq:contstrans}
 \d\vp=\vepsb\psi+\psibar\veps\,,\quad \d\psi=(\pa_t\vp-W'(\vp))\veps\,,\quad \d\psibar=-\vepsb(\pa_t\vp+W'(\vp))\, ,
\end{equation}
into the integration of a total derivative term
\begin{align}
\d S=-\vepsb&\!\! \int\! dt\; \left[(\pa_t \psi)W'(\vp)+\psi W''(\vp)\pa_t \vp\right]\\
&-\veps\!\! \int\! dt\; \left[(\pa_t \psib)W'(\vp)+\psib W''(\vp)\pa_t \vp\right]\, .
\end{align}
For the invariance of the action appropriate (e.~g.\ periodic) boundary conditions must be assumed.
In addition also the Leibniz (chain) rule is used.
In the off-shell formulation of the theory the same terms appear as variation of the action.\footnote{A contribution $F\vp^n+n\psib\vp^{n-1}\psi$ transforms into $(\pa_t\psi)\vp^n+n\psib\vp^{n-1}\pa_t\vp$ since the auxiliary field $F$ generates a derivative of a fermionic field and $\psi$ a derivative of a bosonic field.}

There are similar models in other dimensions (with or without extended supersymmetry).
The most prominent example for this class of models is the four-dimensional ${\cal N}=1$ Wess-Zumino model.
In the off-shell representation it reads (\cite{Wess:1973kz})
\begin{align}
 S=&\int d^4 x\bigg[\half \pa_\mu A \pa^\mu A+\half \pa_\mu B\pa^\mu B+\half \psib \slashed{\pa}\psi-\half F^2-\half G^2\nonumber\\
&+m\left(FA+GB-\half \psib\psi\right)\nonumber\\
&+g\left(FA^2-FB^2+2GAB-\psib(A-i\gamma_5)\psi\right)\bigg]\, .
\label{eq:FourDim}
\end{align}
This model contains the scalar field $A$, the pseudoscalar field $B$, a Majorana-spinor $\psi$ and the auxiliary fields $F$ and $G$.\footnote{The problems of the correct definition of Majorana spinors in Euclidian space and on the lattice will not be discussed here.} 
An important feature of supersymmetric theories is the cancellation between bosonic and fermionic quantum correction.
Bosonic loop diagrams are canceled by fermionic ones in perturbation theory.
In the four-dimensional case this leads to the well-known nonrenormalisation theorem: To get finite expressions only a wave function renormalization ($Z$) that is the same for all fields is needed. The corresponding counterterm is
\begin{equation}
 S_{\text{counter}}=\int d^4 x(Z-1)\bigg[\half \pa_\mu A \pa^\mu A+\half \pa_\mu B\pa^\mu B+\half \psib \slashed{\pa}\psi+\half F^2+\half G^2\bigg].
\end{equation}
As in the one-dimensional case the action \eq{eq:FourDim} is invariant up to terms that can be identified with surface terms when the Leibniz rule is applied.

More generally one can show that a nontrivial Lagrangian can be invariant under all supersymmetry transformations only up to a total derivative term. 
This is a direct consequence of the algebra: The commutator of some supersymmetry transformations applied to a field leads to a derivative of the field. 
All the transformation of a field can hence only be zero if it is constant.
Schematically the action of a more general model can be written as
\begin{align}
\label{eq:Son}
\Son=\inte d^Dx \left( \half \c(-\pa_\mu\pa^\mu+m^2\right. &)\c +\psib(\slashed{\pa}+m) \psi\nonumber \\
&\left.\phantom{\half} +m {\cal V}_1(\c)+{\cal V}_2(\c)+\psib{\cal V}_Y(\c)\psi\right) ,
\end{align}
The action is written here with the interaction vertices $m{\cal V}_1(\c)$ that carry in contrast to ${\cal V}_2(\c)$ besides dependence on the coupling constants also the mass parameter. The vertices of the Yukawa type interaction are summarized by ${\cal V}_Y(\c)$. These vertices can involve a nontrivial dependence in the spinor indices. The $\c$ now can stand for several bosonic (e.~g.\ scalar and pseudoscalar) fields.\footnote{The condensed notation means $\c(-\pa_\mu\pa^\mu+m^2)\c=\sum_r \c_r(-\pa_\mu\pa^\mu+m^2)\c_r$ with, e.~g., in the four dimensions ${\cal N}=1$ model the scalar field $\c_1=A$ and the pseudoscalar field $\c_2=B$.}

Supersymmetry transformations of such a general model lead to a variation of the action consisting of terms of the following form
\begin{equation}
\label{eq:leibinv}
\d S\propto\!\veps\!\! \int\!\! d^\D x  \left[(\pa_\mu \p^{(1)})\cdots\p^{(\nf)}+ \p^{(1)}(\pa_\mu\p^{(2)})\cdots\p^{(\nf)}+\ldots+\p^{(1)}\cdots(\pa_\mu\p^{(\nf)})\right] .
\end{equation}
$\p^{(1)}\cdots \p^{(\nf)}$ denotes a product of fields, of which some may be of the same kind.
One of these fields must be fermionic.
As in the one-dimensional case each of these contributions vanishes if the Leibniz rule holds and appropriate boundary conditions are applied. 
The variation of the off-shell action are of a similar form. 
Just as in one dimension one can easily show in the off-shell formulation that \eq{eq:leibinv} comes from the variation of the
 terms with a product of $\nf$ fields in the action.

The appearance of these contributions can also be understood from a possible superspace representation of the theory. 
The supercharges contain Graßmann and space-time derivatives ($Q\propto\pa_\t+\Theta^\mu \pa_\mu$ with $\Theta^\mu$ depending on the Graßmann coordinates $\theta_\a$, and $\pa_\t$ represents a derivative operator with respect to some of these coordinates). 
The relevant part of the action can be represented by a product of superfields ($\P^{(i)}(x,\t_1,\t_2,\ldots)$) integrated over superspace. 
Hence the supercharges generate the following variation of the action\footnote{Note that is written rather schematically in general one would have to take into account possible constraints that are applied to the superfields and several different supercharges.}
\begin{equation}
\d S\propto \!\!\int \prod_\a d\t_\a d^\D x \left[(\veps Q\Phi^{(1)})\Phi^{(2)}\cdots\Phi^{(\nf)}+\Phi^{(1)}(\veps Q\Phi^{(2)})\cdots\Phi^{(\nf)}+\ldots\right]\, .
\end{equation}
The contributions of the Grassmann derivatives in the supercharge vanishes after the Grassmann integrations.
The above expression \eq{eq:leibinv} can be identified with a component form of this variation.\footnote{In other words: The variation of the highest component (highest power of Grassmann coordinates) of a product of superfields transforms into terms of the form presented in \eq{eq:leibinv} as found in \cite{Dondi:1976tx}.}

This brief review of continuum supersymmetry indicates that the generic form of the variation of the action is \eq{eq:leibinv}.
If the action should be invariant on the lattice as well as in the continuum we have to impose, as can always be done, periodic boundary conditions on the lattice.
The second, more significant, condition for supersymmetry on the lattice is the Leibniz rule.
\section{The failure of the Leibniz rule and supersymmetry breaking on the lattice}
\label{secFailLeibn}
The crucial source of supersymmetry breaking on the lattice is the failure of the Leibniz rule for any discretised derivative operator.
I will discuss possible solutions of this problem in this section. 
In the present discussion of the Leibniz rule the following assumptions are made:
The same lattice derivative operator $\n^\mu$ replaces the continuum derivative in the fermionic and bosonic kinetic part of the action as well as in the supersymmetry transformations.
As a starting point, the lattice action is chosen to be the same as the continuum action \eq{eq:Son} except for a summation instead of the integration and the lattice derivative operator replacing the continuum derivative (the conventions of the lattice formulation are summarized in appendix \ref{sec:Conv}).
This derivative operator is assumed to be antisymmetric $\n^\mu=-(\n^\mu)^T$ and translational invariant.
Its antisymmetric form ensures the hermiticity of the fermionic part of the action.
The doubling problem of such an operator is discussed in section \ref{sec:doubling}, since it introduces an additional source for supersymmetry breaking on the lattice.
In that way, I want to separate the effect of the Leibniz rule from the fermion doubling problem.

As detailed in the last section the Leibniz rule is needed for the invariance of the continuum lattice action.
When the continuum derivative is replaced by a discrete derivative operator the Leibniz rule becomes
\begin{equation}
\label{eq:latleib2}
 (\nabla^\mu(\one{\p}\two{\p}))_m= \one{\p}_m (\nabla^\mu \two{\p})_m+(\nabla^\mu \one{\p})_m\two{\p}_m \, , 
\end{equation}
for arbitrary lattice fields $\one{\p}$ and $\two{\p}$.
One can easily verify the following statement.
\begin{lemma}
 The Leibniz rule, as presented in equation \eq{eq:latleib2}, is violated by all lattice derivative operators.\footnote{To proof of this statement just insert a lattice field that is zero everywhere except on a single lattice point for $\one{\p}$ and $\two{\p}$. I thank A. Wipf for this simple argument.}
\end{lemma}
Although this relation does not hold for any lattice derivative operator, it is still valid in the zero momentum sector, i.~e.\ after the summation of $m$.\footnote{If one splits the antisymmetric matrix $\nabla$ according to \eq{eq:asymrep} 
one can find the modified Leibniz rule for each of the components,
 $\sum_n \nabla^{(r)}_{mn}(\one{\vp}_{n+r}\two{\vp}_n)=(\sum_n \nabla^{(r)}_{mn}\one{\vp}_{n+r})\two{\vp}_{m+r}+\one{\vp}_m(\sum_n \nabla^{(r)}_{mn}\two{\vp}_n)$. Thus the Leibniz rule is fulfilled up to translations, which are not relevant under the summation.}

Since the same derivative operator is used in the transformations the variation of the lattice action is just the 
lattice counterpart of \eq{eq:leibinv}, namely
\begin{multline}
\label{eq:leibinvLatt}
\d S_{\text{L}}\propto \veps \sum_n \left[(\n_\mu \p^{(1)})_n\cdots\p^{(\nf)}_n+\right.\\\left. \p^{(1)}_n(\n_\mu\p^{(2)})_n\cdots\p^{(\nf)}_n+\ldots+\p^{(1)}_n\cdots(\n_\mu\p^{(\nf)})_n\right] .
\end{multline}
The fulfillment of the Leibniz rule after the summations ensures hence the invariance of a quadratic lattice action ($\nf=2$).
In the higher than quadratic case the variation of the action \eq{eq:leibinvLatt} is non-zero as for a product of three ore more fields the analog of \eq{eq:latleib2} is violated even after the summation.
To show this in detail, let us consider such a product of fields in momentum space. The repeated application of \eq{eq:latleib2} 
results in\footnote{Cf.\ appendix \ref{sec:Conv} for the transformation to momentum space.}
\begin{multline}
 (\nabla^\mu(\one{\p}\cdots\nth{\p}))(p_s)=\\ \sum_{k_1\ldots k_{\nf}}\d_L(p_{k_1}+\ldots+p_{k_{\nf}}-p_s)(\nabla^\mu(p_{k_1})+\ldots+\nabla^\mu(p_{k_{\nf}}))\one{\p}\cdots\nth{\p}\, . 
\end{multline}
For the invariance of the action this equation must hold for $p_s=0$ and all fields $\p^{(i)}$.
This is equivalent to \footnote{In momentum space $\nabla^\mu_{mn}$ is periodically continued for momenta larger than the lattice cutoff (cf.\ section \ref{sec:Conv}).} 
\begin{equation}
\label{eq:LeibnizMomentum}
 \d_L(p_{k_1}+\ldots+p_{k_{\nf}})(\nabla^\mu(p_{k_1})+\ldots+\nabla^\mu(p_{k_{\nf}}))=\sum_{i=1}^{\nf-1}\nabla^\mu(p_{k_i})-\nabla^\mu(\sum_{i=1}^{\nf-1}p_{k_i})=0\, .
\end{equation}

The locality of the action is analyzed in the thermodynamic limit, where the discrete momentum becomes continuous, $p_{k_i}\rightarrow p_i$.

For convenience the modulus of $\sum_{i=1}^{\nf-1}p^\mu_{i}$ is first assumed to be smaller than the lattice cutoff $\Lambda^\mu_L=\frac{\pi}{a_\mu}$ for all directions $\mu$. 
These assumptions fix the solution of equation \eq{eq:LeibnizMomentum}  to\footnote{$c_1=1$ in the continuum limit; additional constant contributions are zero because $\nabla^\mu$ is antisymmetric.}
\begin{equation}
\label{eq:SlacandConstants}
 \nabla^\mu(p)=c_1 p^\mu\, .
\end{equation}
This derivative operator is (apart from irrelevant constants) the SLAC-derivative.
The only possible solution is hence known to be nonlocal. 
This is the one of the basic observations of \cite{Kato:2008fr}. 
It does, however, not mean that the Leibniz rule \eq{eq:latleib2} is fulfilled for such a derivative operator. 
The solution is no longer valid, when the modulus of $\sum_{i=1}^{\nf-1}p^\mu_{i}$ becomes larger than the lattice cutoff.
The periodic continuation and the antisymmetry of $\n^\mu$ in momentum space leads to a violation of \eq{eq:LeibnizMomentum}, when the argument of $\n^\mu(p)$ is outside the first Brillouin zone.
The violation is for  $(2l^\mu-1)\Lambda^\mu_L< \sum_{i=1}^{\nf-1}p^\mu_{k_i}<(2l^\mu+1)\Lambda^\mu_L$  ($\forall\mu$; $l^\mu\in \mathbb{Z}$) given by
\begin{equation}
 \sum_{i=1}^{\nf-1}p_{k_i}-\sum_{i=1}^{\nf-1}\nabla(p_{k_i})=2\Lambda_L\sum_\mu l^\mu\, .
\end{equation}
Even a nonlocal derivative operator is hence not enough to ensure the invariance of the lattice action. 
The basic reason of the violation of the Leibniz rule by the SLAC derivative is an incompatibility with the standard multiplication of the algebra of lattice fields.
In this kind of multiplication a momentum of a product of fields, that exceeds the lattice cutoff is mapped back onto the region below this cutoff. 
It induces a periodic lattice ``delta function'' in momentum space (cf.\ \eq{eq:periodicDelta}), which is resolved by a periodic continuation of $\n^\mu(p)$.

One can try to avoid this nonlocality of the lattice action with the introduction of a modified interaction term. 
For example, the product of three fields is represented on the lattice according to 
\begin{equation}
 \int\!\! d^\D\! x\, \one{\p}(x)\two{\p}(x)\three{\p}(x)\,\stackrel{\text{on the lattice}}{\longrightarrow}\, \sum_{m_1,m_2,m_3}\!\! \Ct_{m_1, m_2, m_3}\one{\p}_{m_1}\two{\p}_{m_2}\three{\p}_{m_3}\, .
\label{eq:threeProdrep}
\end{equation}
In Fourier space this ansatz brings equation \eq{eq:LeibnizMomentum} into the form (cf.\ \eq{eq:repFSCt})
\begin{equation}
 \Ct(p_1,p_2,p_3)(\nabla^\mu(p_1)+\nabla^\mu(p_2)+\nabla^\mu(p_3))=0\quad \forall \mu\, ,
\end{equation}
which is solved by 
\begin{equation}
\label{eq:NDSol}
 \Ct(p_1,p_2,p_3)=\prod_\mu\d(\nabla^\mu(p_1)+\nabla^\mu(p_2)+\nabla^\mu(p_3))\, .
\end{equation}
This modification corresponds to a modified product rule on the lattice. 
This kind of product is defined in such a way that after the summation of the lattice index
 $\prod_\mu\d(\sum_{i=1}^{\nf-1}\n^\mu(p_i))$ instead of the usual (periodic) lattice delta $\d_L(\sum_{i=1}^{\nf-1}p_i)$ appears in momentum space.
From the mathematical point of view 
the mentioned incompatibility of the Leibniz rule with the usual product on the lattice suggests such a modification of the product.

For the symmetric derivative $\n^\mu=(\symder)^\mu$ such a solution was proposed in \cite{Dondi:1976tx}.
However, with a derivative different from \eq{eq:SlacandConstants} the solution breaks for continuous momenta the translational invariance on the lattice.
A possible way out is to accept a modification of this invariance \cite{Nojiri:1984ys}. 
Here this property is, however, considered to be even more important than locality.
Another problem of the solution \eq{eq:NDSol} with a symmetric derivative appears when discrete momenta of a finite lattice instead of the continuous momenta in the thermodynamic limit are considered. 
Then it corresponds to a projection onto the trivial solutions, $p_{k_1}=0$, $p_{k_2}=0$, or $p_{k_3}=0$ (as already mentioned in \cite{Dondi:1976tx}).
Consequently, the nonlocal interaction term allows for every finite lattice only a trivial interaction with at least one field at zero momentum, i.~e. 
\begin{multline}
\label{eq:NicolaiDondi}
 \int\!\! d^\D\! x\, \one{\p}(x)\two{\p}(x)\three{\p}(x)\,\stackrel{\text{on the lattice}}{\longrightarrow}\,\sum_{k}\left[ \one{\p}(0)\two{\p}(-p_k)\three{\p}(p_k)\right.\\
\left.+\one{\p}(-p_k)\two{\p}(0)\three{\p}(p_k)+\one{\p}(-p_k)\two{\p}(p_k)\three{\p}(0)\right]\, . 
\end{multline}
It is obvious that such an expression is supersymmetric.
The fields with zero momentum arguments are similar to a mass term for the other fields in the product.
Although for a continuous momentum the correct classical (tree level) continuum limit has been shown in \cite{Dondi:1976tx}, such a solution can not be useful for the lattice simulations.\footnote{Perhaps this interaction term can, however, be interpreted as the interaction with a kind of mean field.}
The nonlocal interaction term alone does not provide a sufficient lattice realization of the theory.

A translational invariant choice for a modified lattice product in \eq{eq:threeProdrep} is $\Ct_{n_1, n_2, n_3}=C_{(n_1-n_2),(n_1-n_3)}$ (cf.\ \eq{eq:repFSC}). 
The resulting product of three fields then leads to 
\begin{multline*}
 \sum_{m_1,m_2,m_3} C_{m_3-m_1,m_3-m_2}\left[(\nabla^\mu\three{\p})_{m_3}\one{\p}_{m_1}\two{\p}_{m_2}+\three{\p}_{m_3}(\nabla^\mu\one{\p})_{m_1}\two{\p}_{m_{2}}+\three{\p}_{m_3}\one{\p}_{m_1}(\nabla^\mu\two{\p})_{m_{2}} \right]\\
=\int_{p_1,p_2, p_3}\three{\p}(-p_1-p_2)C(p_1,p_2)(\nabla(p_1)+\nabla(p_2)-\nabla(p_1+p_2))\one{\p}(p_1)\two{\p}(p_{2})\, ,
\end{multline*}
for the supersymmetry variation of the action.
Obviously, this approach can be generalized to the situation of $\nf$ fields, and the condition for the invariance of the action, cf.\ \eq{eq:LeibnizMomentum}, becomes
\begin{equation}
\label{eq:modLeibnizMomentum} 
C(p_1,\ldots,p_{\nf-1})\left[\sum_{i=1}^{\nf-1}\nabla(p_{i})-\nabla(\sum_{i=1}^{\nf-1}p_{i})\right]=0\, .
\end{equation}
It corresponds to the Leibniz rule for $\nf-1$ fields with the modified lattice product 
\begin{align}
(\one{\p}\ast\two{\p}\ast\cdots\ast\p^{(\nf-1)})_l:&=\sum_{m_1,\ldots , m_{\nf-1}}C_{l-m_1,\ldots , l-m_{\nf-1}}\one{\p}_{m_1}\cdots \p^{(\nf-1)}_{m_{\nf-1}}\nonumber\\
&=\sum_{m_1,\ldots , m_{\nf-1}}\Ct_{l,m_1,\ldots , m_{\nf-1}}\one{\p}_{m_1}\cdots \p^{(\nf-1)}_{m_{\nf-1}}\, .\label{eq:ModProd} 
\end{align}
This time the modification of the product is defined such that after the summation of the lattice index
 the usual (periodic) lattice delta $\quad\d_L(\sum_{i=1}^{\nf-1}p_i)$ is combined with an additional factor $C(p_1,\ldots,p_{\nf})$ in momentum space.
The modified product is commutative as $C$ is assumed to be symmetric under the exchange of its arguments.

With the help of the modified product on the lattice, we arrive at a more general realization of a supersymmetric lattice action.
Although the Leibiz rule \eq{eq:latleib2} can not be fulfilled by any lattice derivative operator the generalized ansatz allows the construction of a supersymmetric realization of a nontrivial lattice action, as detailed later on.
Before I present a concrete realization of such a supersymmetric lattice action, I first address the question in as much locality can be realized in such an approach.
In \cite{Kato:2008fr} a ``No-Go theorem'' for a local supersymmetric lattice realization was found.
Since the effect of the violation of the Leibniz rule by the SLAC derivative for large momenta was not included in these investigations, I get here an even stronger statement for the violation of locality by supersymmetric lattice actions.

\section{A No-Go theorem}
\label{sec:NoGo}
With the results of the discussion presented in the last section I now introduce the assumptions for of a simple No-Go statement.

As above the same antisymmetric lattice derivative operator is assumed to replace the continuum derivative in the action and the supersymmetry transformations. 
In addition to the lattice derivative operator a modified translational invariant product on the lattice \eq{eq:ModProd} is included.
The modification of the product is assumed to appear in every term in the action, except for the kinetic part that contains the terms with the derivatives and the squares of the auxiliary fields. 
It appears also in the nonlinear part of the supersymmetry transformations. 

A kind of ``smoothness condition'' for the limit of infinitely many lattice points is assumed.
This condition should exclude trivial solution similar to \eq{eq:NicolaiDondi}.
 When the number of lattice points increases, also the number of modes ($\p^{(\nf)}(p_{k_{\nf}})$) that are allowed to interact with all modes of the other fields satisfying the constraint $p_{k_{\nf}}+\sum_i^{\nf-1}p_{k_i}=0$ should increase.
Otherwise it is hard to extrapolate the continuum limit, where no other constraint besides $p_{\nf}+\sum_i^{\nf-1}p_i=0$ is imposed on the interacting lattice fields.
In the thermodynamic limit at least in a certain interval around $p_{\nf}=0$ all momenta should be allowed to interact as in the continuum without any further constraint.
As the lattice spacing tends towards zero this interval has to increase.
 
The locality of a lattice derivative operator $(\n^\mu)_{nm}$ means that it decays exponentially when the distance between the lattice points $n$ and $m$ increases.
The width of this exponential decay is proportional to the lattice spacing.
A generalization leads to the locality condition for $C$.
In this case locality means that the elements of $C_{l-m_1,\ldots,l-m_{\nf}}$ decrease exponentially when the distance between the lattice point $l$ and one of the $m_1,\ldots,m_{\nf}$ increases (at fixed distance between $l$ and all the other $m_i$). 

The careful definition of the conditions for a suitable lattice realization is the crucial part of the derivation of the following No-Go statement. Its proof is rather trivial. 

\begin{lemma}[No-Go theorem]
In order to get an interacting supersymmetric lattice theory one needs a nonlocal derivative operator and a nonlocal interaction term.
\end{lemma}
The basic message of this statement is that either locality or supersymmetry must be given up on the lattice.

For the proof of this statement I first recall some well-known facts that relate properties of a function with its Fourier transformed.
If a function is differentiable to all orders, its Fourier transform decays faster than any polynomial.
For an exponential decay it must be analytic in a region around the real axis.
For the Fourier transformed of the lattice operators (in thermodynamic limit) these conditions must not only hold for the whole Brillouin zone.
In addition they must be fulfilled at the boundary of the Brillouin zone, when the function is periodically continued.
Hence locality means nothing but the analyticity of the periodically continued functions $\n^\mu(p)$ and $C(p_1,\ldots,p_{\nf-1})$. 
In the case of $C(p_1,\ldots,p_{\nf-1})$ this must hold with respect to each of the $p_1,\ldots,p_{\nf-1}$ holding all other arguments fixed at an arbitrary value. 
If an analytic function is constant in a certain region (and not only on isolated points) it is constant everywhere.
The same is true for the derivative of an analytic function.
This implies that if the derivative operator has the momentum space representation \eq{eq:SlacandConstants} in a certain region of the BZ, it has to obey this momentum space relation in the complete BZ. 
Otherwise it would violate analyticity inside the BZ.
It is hence a  SLAC type derivative, that is analytic inside the BZ but nonlocal because periodicity is not fulfilled.

The invariance of the action under supersymmetry transformations demands that equation \eq{eq:modLeibnizMomentum} is fulfilled in the interacting case, i.~e.\ $\nf>2$.
Therefore either the term in square brackets or the function $C(p_1,\ldots,p_{\nf-1})$ has to be zero.
According to the smoothness condition there must always be a continuous region for the sum of the $\nf-1$ momenta
where $C(p_1,\ldots,p_{\nf-1})$ is nonzero.
In this region the term in the square bracket must be zero and the derivative has to be \eq{eq:SlacandConstants}.
A derivative operator that has this kind of momentum space representation in a certain region of the BZ can, however, not be analytic and periodic.
Thus the derivative operator must be nonlocal.

On the other hand, the term in square brackets can only be zero in the region where the sum of the $\nf-1$ momenta stays inside the first BZ.
In this case the derivative operator is the nonlocal SLAC derivative.
Hence at least in the region where the sum gets larger than the lattice cutoff $C(p_1,\ldots,p_{\nf-1})$ must be zero.
Since it is zero in a certain region of its arguments it can not be analytic.
Thus also the modified lattice product must be nonlocal.

In contrast to  \cite{Kato:2008fr} the smoothness condition and the effect of the periodic lattic delta are considered here.
This implies that neither a local derivative operator nor a nonlocal product alone can provide a sufficient supersymmetric lattice realization.

A possible approaches to circumvent this No-Go statement is a fine tuning of the bare parameters towards the supersymmetric continuum limit.
Instead of the complete continuum supersymmetry one can also realize only a part of the symmetry on the lattice.
It might be that this reduces the necessary fine tuning, but it can lead to other problems (cf.\ \cite{Kastner:2008zc}).
Like the Ginsparg-Wilson relation for chiral symmetry, a similar relation could lead to a solution for the lattice realization of the symmetry. 
Such a symmetry relation has been found for supersymmetry in \cite{Bergner:2008ws}. 
However, a solution of this relation in terms of an interacting local lattice action was so far not constructed.
In \cite{Kato:2008fr} a certain realization using matrices instead of fields was proposed.
Although it allows to circumvent the No-Go statement it induces an infinite number of degrees of freedom in the continuum limit.

Different from these attempts I will construct here a nonlocal lattice representation.
This allows for a way to find a supersymmetric lattice realization.
Instead of supersymmetry one can investigate how locality is restored in the continuum limit.
At least in low dimensions no fine tuning is necessary for this restoration.
I will show in perturbative calculations and numerical simulations that correct results can be obtained from such a nonlocal lattice theory.
\section{A realization of the complete supersymmetric on the lattice}
\label{sec:Real}
In this section I will construct a lattice action that preserves all supersymmetries.
A similar lattice realization was constructed in \cite{Bartels:1982ue} in the thermodynamic limit starting from a momentum space representation of the theory.
Here I will formulate the lattice theory in a different way, that allows for a computation in the simulations.

I start with the off-shell representation of the theory.
Since anyway a nonlocal lattice derivative is needed one can, for convenience, choose the SLAC derivative ($\n^\mu=(\derSLAC)^\mu$) in the lattice action and the transformations.
The translation of a product of continuum fields into a nonlocal lattice term, cf.\ \eq{eq:ModProd}, is done for each term of the action separately according to the number $\nf$ of fields appearing in the product. 
A quadratic term needs no further modification.
For a number of $\nf>2$ fields (no matter if they are bosonic, fermionic, or auxiliary) the appropriate $C(p_{k_1},p_{k_2},\ldots,p_{k_{\nf-1}})$ must be included.
The $C$s  can be decomposed into a product of their one-dimensional counterparts $C_1$,
\begin{equation}
 C(p_{k_1},p_{k_2},\ldots,p_{k_{\nf-1}})=\prod_\mu C_1(p^\mu_{k_1},p^\mu_2,\ldots,p^\mu_{k_{\nf-1}})\, .
\end{equation}
As found above in a realization with the SLAC derivative the nonlocal product has to exclude all momenta with a sum that exceeds the lattice cutoff. 
The following  $C$ solves \eq{eq:modLeibnizMomentum} and leads to a supersymmetric lattice action:
\begin{align}
C_1(p^\mu_{k_1},p^\mu_{k_2},\ldots,p^\mu_{k_{\nf-1}}):=
\begin{cases}
 0 & \text{if  } |\sum_{i=1}^{\nf-1} p_{k_i}|>\Lambda^\mu_L \\
1 &\text{otherwise }
\end{cases}\, .
\end{align}
In such a way a nonperiodic delta is introduced in the lattice theory,
\begin{equation}
\label{eq:nonperdelta}
 \d^{(\nf-1)}(p_{k_1}+p_{k_2}+\ldots+p_{k_{\nf}}):=\d_L(p_{k_1}+\ldots+p_{k_{\nf}})C(p_{k_1},p_{k_2},\ldots,p_{k_{\nf-1}})\, .
\end{equation}
This nonperiodic delta function is nonzero only if the sum of the $\nf-1$ momenta is equal to $p_{k_{\nf}}$.
A nonperiodic delta was also introduced in \cite{Bartels:1982ue}.

For the purpose of the lattice simulation, an appropriate representation of the modified product with a nonperiodic delta is needed.
Since $C$ can in higher dimension be represented as a product of its one-dimensional counterparts, it is enough to find a solution the one-dimensional case.
Consider a lattice with the $(\nf-1)$-fold number of lattice points compared to the original lattice.
The lattice spacing of this finer lattice is $a/(\nf-1)$.
Hence the periodicity of a Fourier space delta of this lattice ($\d^{\nf-1}$) is $(\nf-1)\Lambda_L$. 
The boundary of the first BZ of this finer lattice cannot be reached with the sum of the momenta $p_{k_1}\ldots p_{k_{\nf-1}}$. They are hence never folded back and the $\d^{\nf-1}(p_{k_1}+\ldots+ p_{k_{\nf}})$  is equal to one only when $\sum_{i=1}^{\nf-1} p_{k_i}$ exactly matches $p_{\nf}$, and otherwise zero.
Thus it represents a nonperiodic delta. 
A one-dimensional interaction term with a nonlocal product of $\nf$ fields can hence be represented on the lattice in the following way (cf.\ equation \eq{eq:FCDef})
\begin{multline}
 \int dx \p^{(1)}(x)\cdots\nth{\p}(x) \stackrel{\text{on the lattice}}{\rightarrow}
 \sum_{k_1,\ldots,k_{\nf}}\!\!\!\d^{\nf-1}(p_{k_1}+\ldots+p_{k_{\nf}}) \p^{(1)}(p_{k_1})\cdots\nth{\p}(p_{k_{\nf}})\\
=\sum_{k_1,\ldots,k_{\nf}}\frac{a}{\nf-1}\sum_{m=0}^{(\nf-1) N-1}e^{-i\frac{a}{\nf-1}m(p_{k_1}+\ldots+p_{k_{\nf}})}\p^{(1)}(p_{k_1})\cdots\nth{\p}(p_{k_{\nf}})\, .
\end{multline}
It is clear that a representation of a nonperiodic delta is also provided when $\d^{m-1}$ with $m>\nf$ instead of $\d^{\nf-1}$ is used.
Consequently for all products $\d^{m-1}$ can be used, where $m$  the highest appearing power of the fields ($\nf$).

Going back from momentum to real space representation one arrives at\footnote{The ordinary definition of the Fourier representation of the fields $\p^{(i)}_{n}$ is employed.}
\begin{multline}
 \int dx \p^{(1)}(x)\cdots\nth{\p}(x) \stackrel{\text{on the lattice}}{\rightarrow}\\ \frac{a}{\nf-1}\sum_{n=0}^{(\nf-1)N-1}\tilde{\p}^{(1)}_n\cdots\nth{\tilde{\p}}_n\;\text{ with }\; \tilde{\p}^{(i)}_n=\sum_m \Fc_{nm}\p^{(i)}_m\, .
\end{multline}
The $\big((\nf-1)N\big)\times N$ matrix $\Fc_{nm}$ translates all of the fields into fields on the finer lattice. It comprises a Fourier transformation on the lattice with $N$ lattice points and an inverse Fourier transformation on the larger lattice with $(\nf-1) N$ lattice points. The matrix elements read explicitly
\begin{equation}
\Fc_{nm}=\frac{\sin(\pi(m-n/(\nf-1))}{aN\sin(\frac{\pi}{N}(m-n/(\nf-1)))}\text{ for }m\neq n/(\nf-1)\;\;\text{ and } 1 \text{ otherwise.}
\label{eq:defFnm}
\end{equation}
In this representation the nonlocal product is defined by
\begin{equation}
\Ct_{n_1,\ldots,n_{\nf}}=\frac{a}{\nf-1}\sum_{n=0}^{(\nf-1)N-1}\Fc_{mn_1}\cdots \Fc_{mn_{\nf}}\, .
\end{equation}

This approach can easily be generalized to the higher-dimensional case.
One obtains a fully supersymmetric lattice action, which can be used in numerical simulations.
The result of the simulations can be found in the section \ref{sec:Simulations}.

For the on-shell representation of the theory the approach can also be applied.
The modification of the product must then be chosen according to the power of fields that appear in the corresponding variation of the action \eq{eq:leibinv}.
The same modification of the product must also be applied for the nonlinear part of the supersymmetry transformations.

The considered lattice action can also be represented completely on the finer lattice with $\nf-1$ lattice points, if $\nf$ is the largest number of fields that appear in a product.
It leads to a theory where the momentum of all fields is constrained below the cutoff $\Lambda/(\nf-1)$ instead of $\Lambda$.

According to the No-Go theorem there are also other possibilities to realize an intact supersymmetry on the lattice.
In each of these representations the nonlocal product constraints the momentum in the interaction terms to a region, where the Leibniz rule is fulfilled.
For a high enough momenta of the fields there are effectively no interaction terms present in the theory.
The interacting theory hence effectively lives on a lattice with a larger lattice spacing.
The advantage of the approach presented in this section is that a contribution of the noninteracting modes is not present.
\section{The standard discretisation and supersymmetry breaking on the lattice due to the fermion doubling problem}
\label{sec:doubling}
In the last sections I have discussed the violation of supersymmetry due to the absence of the Leibniz rule.
In this discussions I have assumed an antisymmetric derivative operator in the lattice action.
It is well-known that such a derivative operator introduces a doubling problem.
I will discuss the doubling problem here, separately. 
This will also explain the assumption of the same derivative operator for fermions and bosons.
Let us therefore allow different derivative operators and masses  for fermions and bosons.
The continuum derivative operators $\pa_\mu$ in the fermionic and $\pa_\mu\pa^\mu$ in the bosonic sector are replaced by lattice difference operators $\n^\mu_{nm}$ and $\Box_{nm}$. 
The continuum action \eq{eq:Son} is represented on the lattice by
\begin{multline}
\label{eq:SonLattice}
 \Son=\inte d^\D x \left( \half \p(-\Box+\mb^2)\p+\psib(\slashed{\n}_{\text{f}}+\mf) \psi\right.\\\left. +\mb {\cal V}_1(\p)+{\cal V}_2(\p)+\psib{\cal V}_Y(\p)\psi\right) .
\end{multline}
In the simplest discretisation in the bosonic sector is the forward and backward derivative ($\Box=\sum_\mu \rightder_\mu \leftder_\mu$, $(\mb)_{mn}=m\d_{mn}$).
The discretisation in the fermionic sector is more difficult.
It is well-known that a naive discretisation, i.~e.\ with $\symder$, introduces a doubling of the fermion species in the continuum limit.
In accordance with the Nielson-Ninomiya theorem \cite{Nielsen:1981hk,Friedan:1982nk} all local representations with an antisymmetric derivative operator share this problem.
The only way out is an additional contribution that is symmetric in momentum space ($\MW(-p)=\MW(p)$).
The common way to add such a contribution to the action is with the identity in spinor space.
Then it resembles a momentum dependent mass.
To remove the doubling problem, this mass diverges at the additional zero modes  (doublers) in the continuum limit.
The doublers have then no dynamic contribution to the action and are ``freezed out''.
The most prominent example of such a term is the Wilson mass. It is
\begin{equation}
 \MW=\frac{ar}{2}\sum_\mu \rightder_\mu \leftder_\mu\, .
\end{equation}

In the case of supersymmetry the doubling modes are problematic.
In the naive lattice formulation they appear only in the fermionic sector.
Consequently, there are no longer the same number of degrees of freedom in the bosonic and fermionic sector, and supersymmetry is broken.
As described above one can try to resolve this problem with a Wilson-type mass term in the fermionic sector
and set $\mf=m+\MW$.
Due to the additional mass term $\MW$ and the different derivative operators in the bosonic and fermionic sector supersymmetry is, however, violated even in the free theory. 
Classically it is restored in the continuum limit, where the difference between the derivative operators and the additional mass term vanishes.
In the quantum theory divergencies of the loop contributions alter this classical behavior and violate supersymmetry even in the continuum limit.
Thus neither the difference between different derivative operators for fermions and bosons nor the additional Wilson mass term can in principle be assumed to have no influence on the continuum limit.

Indeed, the cancellations of bosonic and fermionic loop contributions are based on the relations between the bosonic and fermionic vertices and propagators.
These relations are, however, not valid in this discretisations, and the continuum cancellation do not appear in lattice perturbation theory.
They can be recovered only in the continuum limit.
To find out the relevance of certain contributions in lattice perturbation theory for the continuum limit, one can use the lattice degree of divergence defined by Reisz \cite{Reisz1988pc}.\footnote{As usual for lattice perturbation theory the calculation is carried out in the thermodynamic limit.}
A one loop diagram in lattice perturbation theory is represented as an integral of the loop momentum $p$ constrained to the first BZ.
When the lattice degree of divergence is negative, the continuum limit can be obtained in a naive way:
The integrand is taken in the limit of $a\rightarrow 0$ and afterwards an unconstrained momentum integration yields the correct continuum limit. 
In such a naive continuum limit the contribution of the Wilson mass and of the difference between the derivative operators disappears (the essential part for this limit comes from the vicinity of $p=0$).
A typical one loop diagram with fermionic and bosonic contributions that cancel in the corresponding continuum expressions is
\begin{align}
\label{eq:propCon}
 &\int \frac{d^\D p}{(2\pi)^\D}\left(\frac{\mb}{-\Box(p)+\mb^2}-\frac{\mf(p)}{-\sum_\mu \symder_\mu(p)\symder_\mu(p)+(\mf(p))^2}\right)\nonumber\\
&\quad=\int \frac{d^\D p}{(2\pi)^\D}\left(\frac{\mb(\Box-\symder_\mu\symder_\mu+\mf^2-\mb^2)}{(-\Box+\mb^2)(-\sum_\mu \symder_\mu\symder_\mu+\mf^2)}\right.\\
&\qquad\qquad\hspace*{4cm}\qquad\left.+\frac{\mb-\mf}{-\sum_\mu \symder_\mu\symder_\mu+\mf^2}\right)\label{eq:WilsonCont}
\, .
\end{align}
The last term \eq{eq:WilsonCont} is nonzero due to the Wilson mass term for the fermions.
The bosonic mass parameter $\mb$ that appear there comes from one of the vertices $m{\cal V}_1$.
The lattice degree of divergence of this term is $\D-1$. 
Already in one dimension the necessary cancellation between fermionic and bosonic loops is hence not recovered in the continuum limit.
This effect was investigated in detail in \cite{Giedt:2004vb} (cf.\ also \cite{Catterall:2000rv,Bergner:2007pu}).
It explains the non-supersymmetric continuum limit of a lattice realization
with a straight forward implementation and a Wilson mass only for the fermions.

The first term \eq{eq:propCon} is due to the difference between the bosonic and fermionic propagators.
Its lattice degree of divergence, $\D-2$, is smaller.
In one dimension it is not relevant, whereas in two dimensions it can lead to a supersymmetry breaking in the continuum.

To avoid this kind of supersymmetry breaking it is necessary to use the same propagators for fermions and bosons ($\Box=(\n_{\text{f}})_\mu(\n_{\text{f}})^\mu$; $\mf=\mb$).  
The derivative operator then introduces a doubling problem also in the bosonic sector that is removed with the same Wilson mass as for the fermions.
It is crucial that this mass parameter appears also in the bosonic vertices of the theory ($\mb{\cal V}_1=\mf{\cal V}_1$).
This nontrivial modification of the theory leads to the correct cancellations of the considered fermionic and bosonic loop diagrams.
The Wilson mass is hence a modification of the usual mass term in the superpotential.
Such kind of lattice formulations were already investigated, e.~g., in \cite{Catterall:2000rv,Golterman1989307}.
However, the additional vertices introduced by the Wilson mass can modify or violate symmetries of the model on the lattice as found in \cite{Kastner:2008zc}.
The SLAC derivative opens the possibility to circumvent this problem without a modification of the superpotential. 
Since it is nonlocal, it does not introduce any doublers, and a Wilson mass term is not needed.

The described adjustment of the fermionic and bosonic mass terms and derivatives alone does not imply a supersymmetric continuum limit.
This can be seen even from the perturbative point of view.
The cancellation of fermionic and bosonic loop diagrams is present, as it is in the continuum.
One can show that the nonrenormalization theorems of the superpotential in the off-shell theory still holds on the lattice as well as in the continuum.
Nevertheless in the four-dimensional model the wave function renormalization is not the same for all fields \cite{Bartels:1982ue}, even in the continuum limit.
In this way the violation of the Leibniz rule is present in the lattice perturbation theory.
Although the masses and derivatives are the same for fermions and bosons non-supersymmetric counterterms are needed to get a supersymmetric continuum limit.
\section{Lattice perturbation theory of the nonlocal lattice theory}
\label{sec:PertTh}
In the last sections I have shown that only lattice realizations with nonlocal operators can be invariant under supersymmetry on the lattice.
An explicit supersymmetric lattice action can be constructed with such operators.
In addition nonlocal operators do not need additional Wilson mass terms.
Although these mass terms can be consistently included in the superpotential they still introduce a serious modification of the on-shell theory.

On the other hand the locality of the continuum theory should be respected in the continuum limit.
For the SLAC derivative in lattice QED Karsten and Smit have shown in \cite{Karsten1979100} that nonlocal and noncovariant counterterms are necessary to achieve this.\footnote{In \cite{Rabin:1981nm} a different conclusion was drawn that was criticized in \cite{Karsten1981103}.}
According to the usual argument this shows that nonlocal lattice actions are not allowed in the simulations.

Let us first consider the SLAC derivative, but no modification of the product in the interaction terms.\footnote{This means $\Box=\n_{\text{f}}^2$, $(\mb)_{nm}=(\mf)_{mn}=m\d_{mn}$, $\n_{\text{f}}=\derSLAC$ in equation \eq {eq:SonLattice}.} 
As has been shown in \cite{Bergner:2007pu}, the twodimensional Wess-Zumino models needs no nonlocal or noncovariant counterterms to achieve the correct continuum limit. 
The propagators of a theory with the SLAC derivative are the same as in the continuum.
The only difference to the continuum perturbation theory is a lattice momentum conservation at each vertex (with a periodic delta) and the restriction of each momentum integration to the first BZ.
As a lattice degree of divergence of the diagrams it is enough to take the degree of divergence of the corresponding continuum diagram.\footnote{The degree of divergence defined in \cite{Reisz1988pc} can, strictly speaking , not be applied here since it assumes that the integrands are periodic smooth functions of the momentum.} 
The nonperiodic delta leads to so called ``umklapp'' contributions in the loop diagrams \cite{Rabin:1981nm,Sharatchandra:1976af}, where the momentum in one of these delta functions becomes larger than the lattice cutoff.
In two dimension these contributions vanish in the continuum limit \cite{Bergner:2007pu}.
For the situation in more than two dimensions it was claimed in \cite{Rabin:1981nm} that that the correct continuum limit is obtained without any nonlocal or non-covariant counterterms and the ``umklapp'' contributions are irrelevant.

The only difference of this model with the SLAC derivative and the full supersymmetric model of section \ref{sec:Real} is the absence of the ``umklapp'' contributions.
All momenta that appear in the propagators are restriction to the first BZ due to the nonlocal product \eq{eq:ModProd}.
Hence, apart from the restrictions of the integration of the loop momenta, the diagrams are the same as in the continuum.
If the ``umklapp'' contributions are irrelevant the continuum limit is obtained with the same counterterms as in the case without a nonlocal interaction.
This is certainly true for the one- and two-dimensional models. 
Recently a more explicit proof of this fact was presented in \cite{Kadoh:2009sp}. 
In addition it was pointed out that the constraints on the momentum integration lead to new additional counterterms in the four-dimensional theory.
It is still not clear in as much these results are consistent with \cite{Rabin:1981nm} and whether no nonlocal counterterms are necessary in the four-dimensional theory.
\section{Simulation of a complete realization of supersymmetry on the lattice in a one-dimensional model}
\label{sec:Simulations}
The aim of the present work is not only to suggest a possible nonlocal realizations, as done in section \ref{sec:Real}.
The solution were also checked on the nonperturbative level.
Therefore lattice simulations of the one-dimensional full supersymmetric model were performed.
In this way I could measure directly signs of a completely realized supersymmetry on the lattice.
The continuum model is chosen to be the same as in \cite{Bergner:2007pu}.
This will allow a direct comparison with the discretisations considered there.
\subsection{The discretised actions}
The starting point is a discretised version of the continuum action \eq{eq:cont1D}
\begin{equation}
\label{eq:latact}
 S_{\text{L}}= \frac{1}{2} \sum_{n}\left((\derSLAC\vp)_n^2 + W'_L(\vp)_n^2\right)+ \sum_{n,m}\psb_n\,\big(\derSLAC_{nm}
+W''_L(\vp)_{nm}\big)\,\psi_m\, .
\end{equation}
In the continuum theory the superpotential is chosen to be $W(\vp)=\frac{m}{2}\vp^2+\frac{g}{4}\vp^4$.
The derivative operator is in all of the discretisations considered here the SLAC derivative.
The results of a discretised version with different implementations of the Wilson mass term can be found in \cite{Bergner:2007pu}. 

On the lattice the following supersymmetry transformations are considered
\begin{align}
\label{eq:varitsep1}
  &\d^{(1)} \vp_n=\vepsb\psi_n\; ;\quad \d^{(1)}\psi_n=0\; ;\quad \d^{(1)} \psib_n=-\vepsb(\derSLAC\vp+W'_L(\vp))_n\\
  &\d^{(2)} \vp_n=\psib_n\veps\; ;\quad \d^{(2)}\psi_n=(\derSLAC\vp-W'_L(\vp))_n\veps\; ;\quad \d^{(1)} \psib_n=0\, .
\label{eq:varitsep2}
\end{align}
The discretisation of the continuum transformations \eq{eq:contstrans} is in this way chosen according to the assumptions of section \ref{secFailLeibn} and \ref{sec:NoGo} with the same derivative operator as in the action.

In the simplest discretisation no modification of the lattice product is included.
Then the interaction is represented by
\begin{equation}
\label{eq:vert1}
W'_L(\vp)_n=m+g \vp_n^{3}\, ,
\end{equation}
and
\begin{equation}
\label{eq:vert2}
W''_L(\vp)_{nm}=(m+3 g\vp_{m}^{2})\d_{nm}\, .
\end{equation}
This discretisation is called here the \emph{unimproved model}.
The action is not invariant under the two supersymmetry transformations.

As in \cite{Bergner:2007pu} a second discretisation of the model is constructed with the so called ``Nicolai improvement''.
The difference between \eq{eq:latact} and the corresponding lattice action are discretisations of surface terms, which are assumed to vanish in the continuum limit.
It reads
\begin{equation}
 S_{\text{LI}}= \frac{1}{2} \sum_{n}\left(\derSLAC\vp + W'_L(\vp)\right)_n^2+ \sum_{n,m}\psb_n\,\big(\derSLAC_{nm}
+W''_L(\vp)_{nm}\big)\,\psi_m\, ,
\end{equation}
with the interaction vertices \eq{eq:vert1} and \eq{eq:vert2} as in the unimproved model.
It is invariant under the supersymmetry transformation \eq{eq:varitsep1}, but not under the second supersymmetry transformation \eq{eq:varitsep2}.
This discretised lattice model is called here the \emph{improved model}.

Finally I present here the first simulations with a model that respects all of the supersymmetry on the lattice.
According to the construction in section \ref{sec:Real} the continuum bosonic interaction vertices are represented on the lattice with
\begin{equation}
W'_L(\vp)_n=m\vp_n+\frac{ag}{3}\sum_{m=0}^{3N-1} \Ft_{mn}(\tilde{\vp}_m)^{3}\, ,
\end{equation}
and the fermionic ones with
\begin{equation}
W''_L(\vp)_{nm}=m\d_{nm}+ag\sum_{m_1=0}^{3N-1} \Ft_{m_1n}\Ft_{m_1m}(\tilde{\vp}_{m_1})^{2}\, .
\end{equation}
For the current choice of the superpotential the fields $\tilde{\vp}$ are
\begin{equation}
 \tilde{\vp}_n=\sum_m\Ft_{nm}\vp_m\, ,
\end{equation}
with $\Ft$ according to \eq{eq:defFnm}.
With this choice for $W'_L$ the difference between the improved and unimproved discretisation vanishes already at finite lattice spacing.
The model is called here the \emph{full supersymmetric model}.
\FIGURE[!ht]{
  \includegraphics[width=\grwidth]{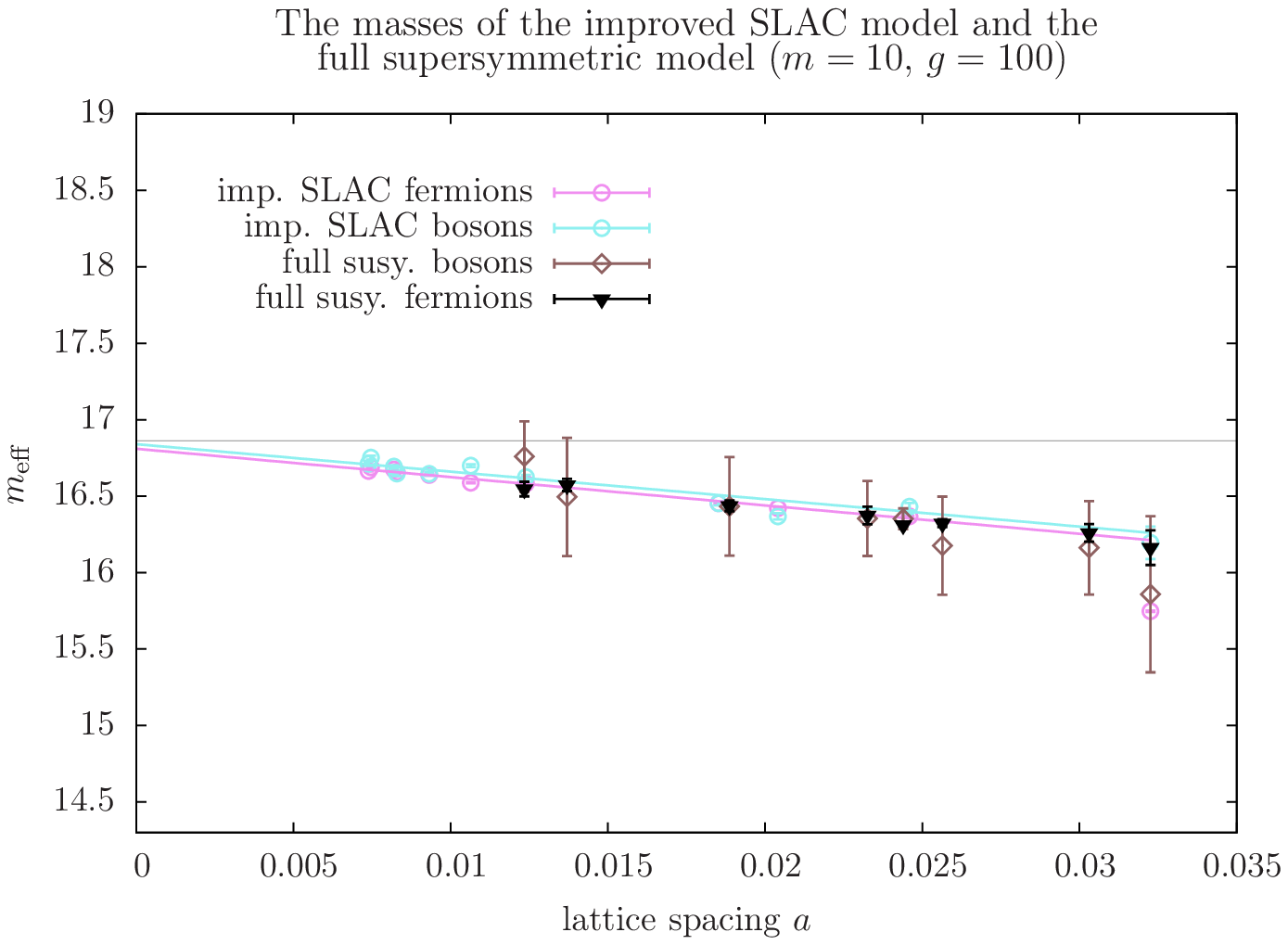}
  \caption{The masses for fermions and bosons of the full supersymmetric model in comparison to the results of the improved model presented in \cite{Bergner:2007pu}. The predicted continuum value is $16.865$. In addition the linear fit of the data of the improved model (fermions: $16.81 -18.53 a$; bosons $16.84-17.97 a$) is shown.}
  \label{fig:Mass}}

\subsection{The algorithm}
For the simulations the same algorithm as used in \cite{Bergner:2007pu} was applied.
Supersymmetry demands for the simulation of dynamical fermions. 
This leads to a contribution of the fermionic determinant in addition to the bosonic part $S_B$ of the action in the path integral measure,
\begin{equation}
 S_{\text{eff}}(\vp)=S_B(\vp)-\log\det K_f(\vp)\, .
\end{equation}
The HMC algorithm \cite{Duane1987216} is applied.
The update algorithm follows a molecular dynamics trajectory determined by
\begin{equation}
 \dot{\vp}_n=\frac{\pa \mathcal{H}}{\pa\pi_n}\, ,\quad \dot{\pi}_n=-\frac{\pa \mathcal{H}}{\pa\vp_n}\, ,
\label{eq:eqmHMC}
\end{equation}
and the Hamiltonian
\begin{equation}
\mathcal{ H}=\half\sum_{n=0}^{N-1}\pi_n+S_{\text{eff}}(\vp)\, .
\end{equation}
The numerical solutions of the differential equations \eq{eq:eqmHMC} are computed with a standard leap frog algorithm.
The fermionic contribution $\pa S_{\text{eff}}(\vp)/\pa \vp_n$ is calculated from
\begin{equation}
\frac{\pa}{\pa\vp_n}  (\log\det K_f)=\frac{\pa}{\pa\vp_n}(\tr\log K_f)=\tr\left(\left(\frac{\pa K_f}{\pa\vp_n}\right) K_f^{-1}\right)\, ,
\label{eq:ForceFerm}
\end{equation}
where
\begin{equation}
 \frac{\pa K_f}{\pa\vp_n}=\frac{\pa W''_L(\vp)_{nm}}{\pa\vp_n}\, .
\end{equation}
When $W''_L$ is diagonal, the trace collapses. This reduces the cost of the numerical calculations with a diagonal interaction term.
This is not the case for the realization with a nonlocal interaction term, where the trace includes a sum over all lattice points.
Nevertheless the simulation is still possible at least in lower dimensions.
\subsection{Numerical results: Masses and Ward-identities}

To get an indication of supersymmetry on the lattice the masses and Ward-identities of the theory are measured on the lattice in the same way as in \cite{Bergner:2007pu}.
The masses are obtained from the exponential decay of the fermionic and bosonic correlators.
The fermionic correlator of the SLAC derivative shows appart from the exponential decay an additional oscillating contribution that vanishes in the continuum limit.
To enhance the signal for the fermionic mass already at finite lattice spacing a filtering technique was applied (for details cf.\ \cite{Bergner:2007pu}).
\FIGURE[!ht]{
  \includegraphics[width=\grwidth]{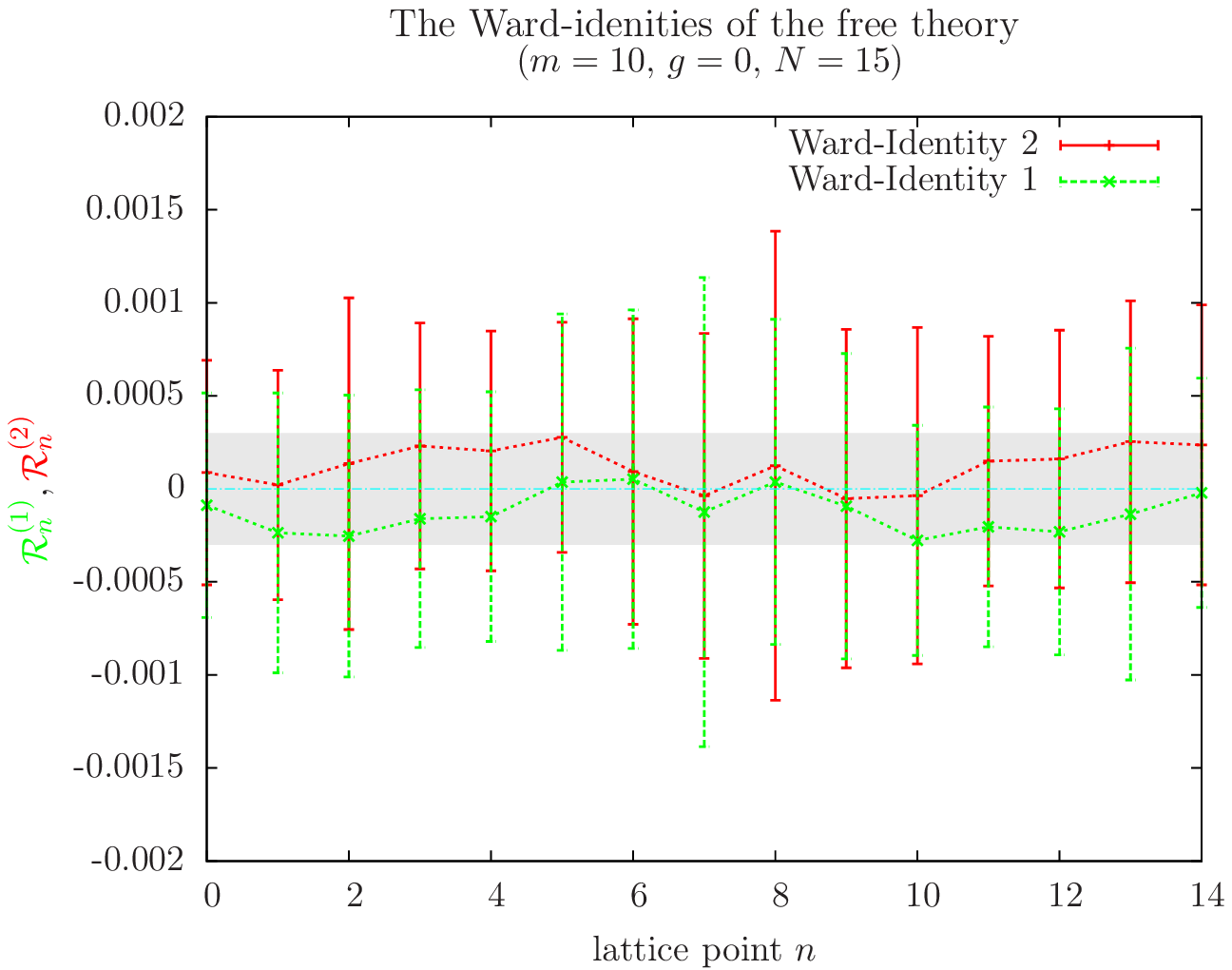}
  \caption{The results of a measurement of the Ward identity in the free theory at small lattice spacing. This identifies the deviation from zero due to the statistical or systematic errors. No larger deviation from zero than $0.0003$ is observed.}
  \label{fig:FT}}
\FIGURE[!ht]{
\includegraphics[width=\grwidth]{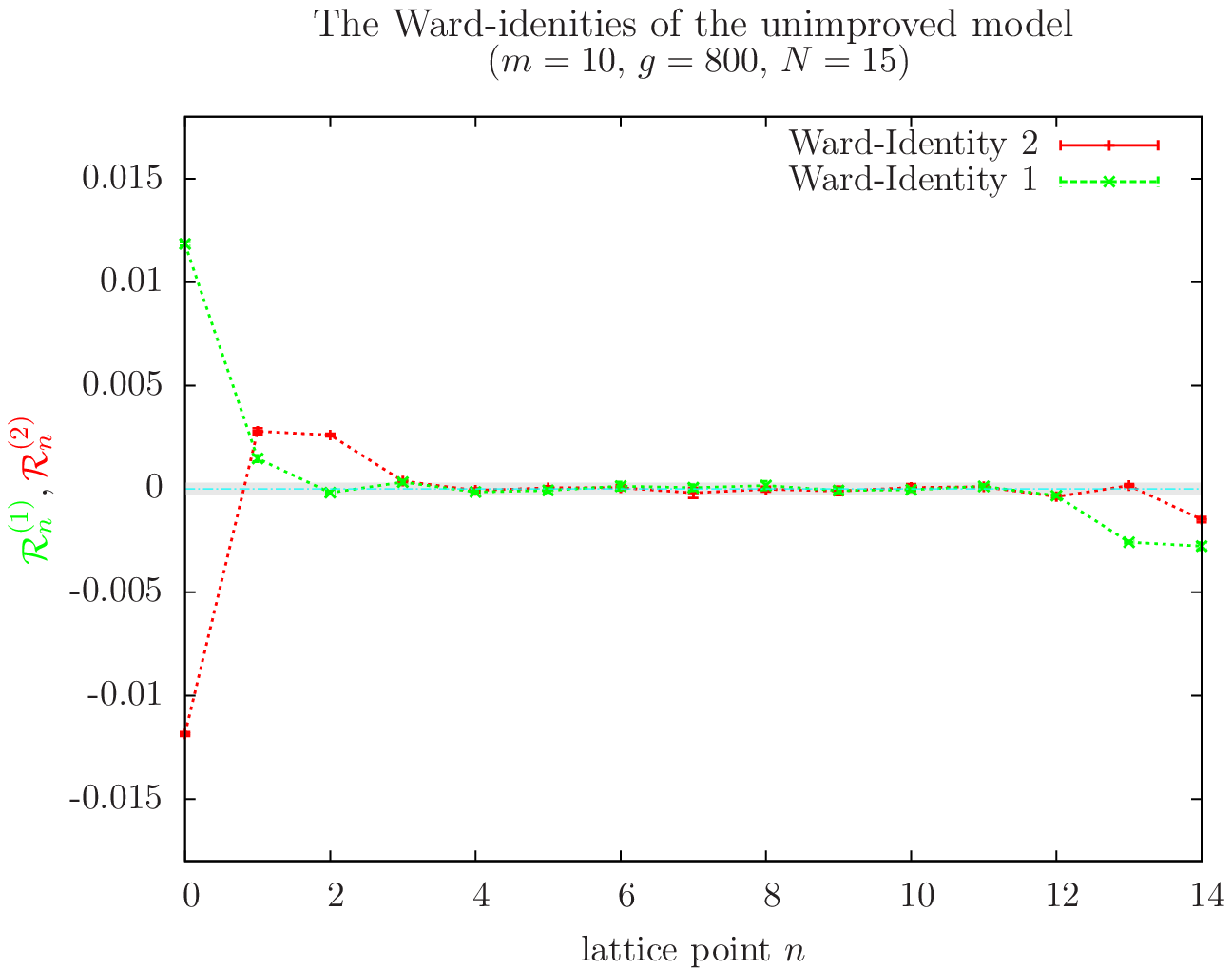}
  \caption{The Ward identity of the unimproved model at a larger coupling. The deviation from zero is clearly above the errorbounds.}
  \label{fig:SLACU}}
\FIGURE[!ht]{
\includegraphics[width=\grwidth]{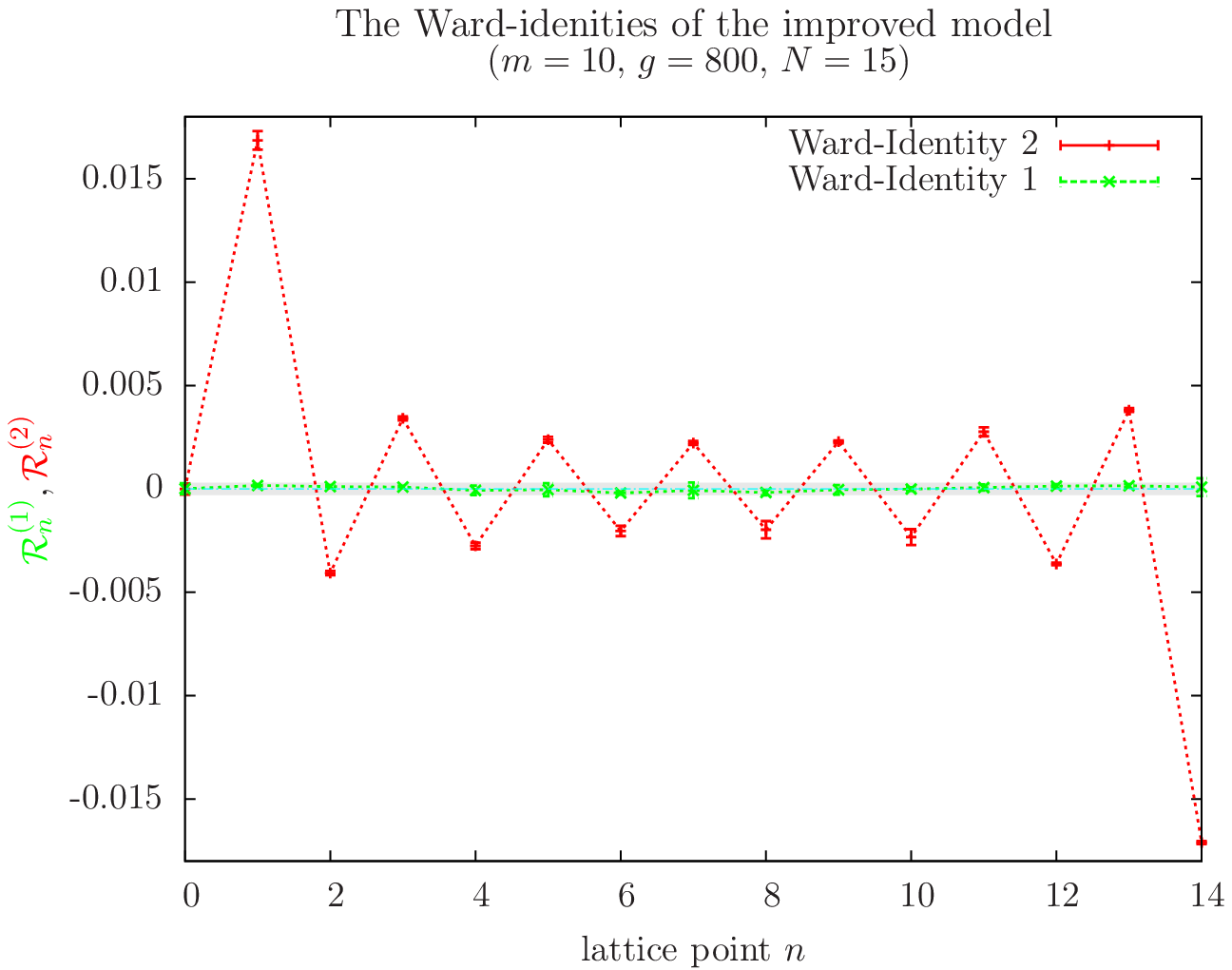}
  \caption{The Ward identity of the improved model. The Ward-identity 1 is zero within the errors. On the other hand Ward identity 2 shows a deviation from zeronearly two orders of magnitude above the errorbounds.}
  \label{fig:SLACI}}
\FIGURE{\includegraphics[width=\grwidth]{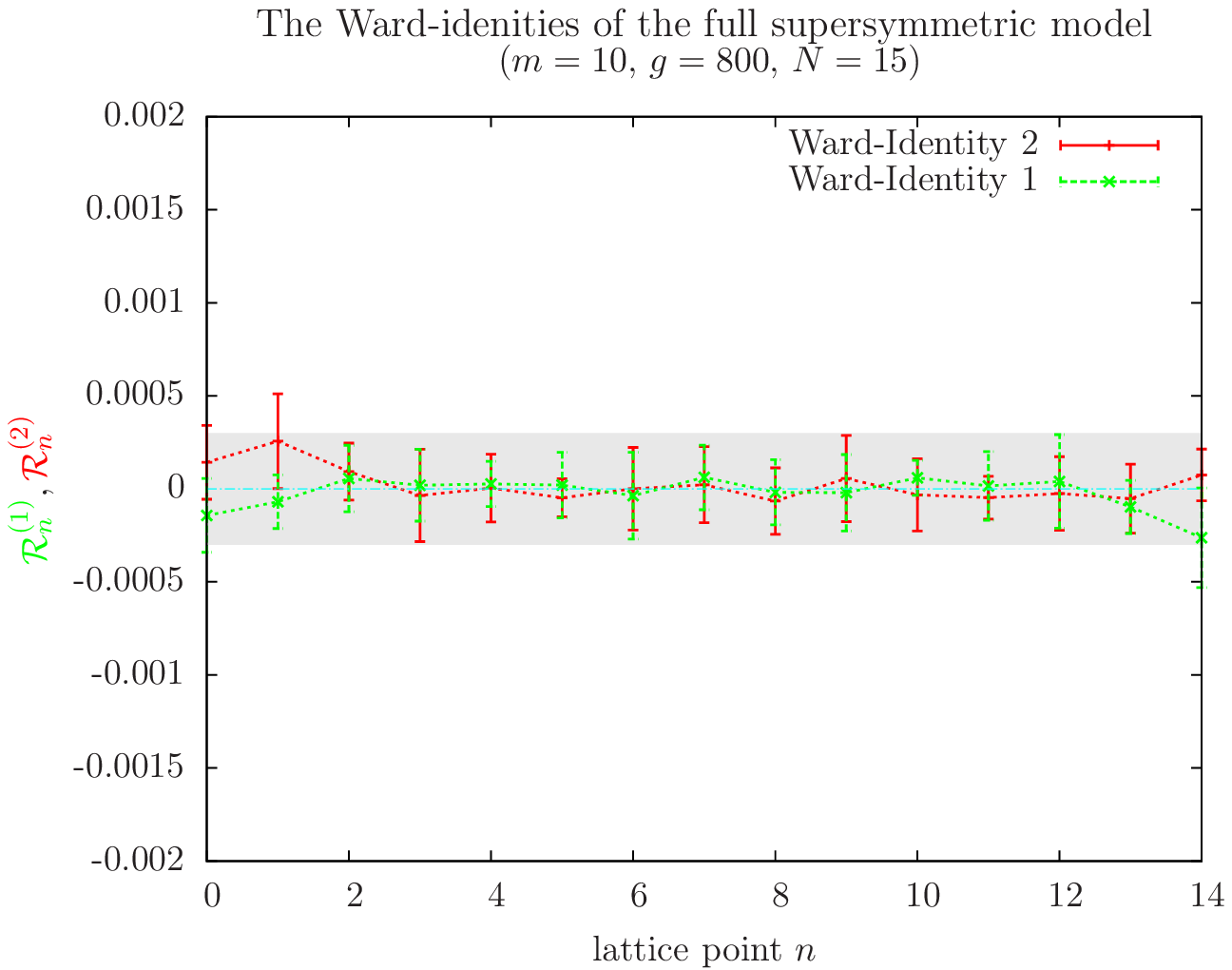}
  \caption{The Ward identity of the full supersymmetric model. Within the determined small errorbounds it is zero.}
  \label{fig:FS}}
In addition the following Ward-Identities were measured
\begin{align}
  \vepsb \mathcal{R}^{(1)}_{n-m}&=\langle\vp_n\,
    \delta^{(1)}\psb_m\rangle + \langle(\delta^{(1)}\vp_n)\psb_m\rangle\nonumber\\
 \text{Ward identity 1:}\quad\mathcal{R}^{(1)}_{n-m}&= \langle\psi_n\psb_m\rangle - \langle\vp_n (\derSLAC\vp)_m\rangle -
  \langle\vp_n W'_L(\vp)_m\rangle\\
 -\mathcal{R}^{(2)}_{n-m}\veps &=\langle\vp_n\,\delta^{(2)}
    \psi_m\rangle + \langle(\delta^{(2)}\vp_n)\psi_m\rangle\nonumber\\
\text{Ward identity 2:}\quad\mathcal{R}^{(2)}_{n-m}& =\langle\psb_n\psi_m\rangle- \langle\vp_n\,(\derSLAC\vp)_m\rangle + \langle\vp_n W'_L(\vp)_m\rangle  \, .
\end{align}
If supersymmetry is realized both of them must be identical zero.

The obtained masses show not much difference for the considered discretisations.
The masses of the improved and unimproved model were already presented in \cite{Bergner:2007pu}.
A good agreement between fermionic and bosonic masses was measured in both models.
In comparison to a discretisation with the Wilson derivative\footnote{This means the symmetric derivative and the Wilson mass term in the fermionic and bosonic sector as discussed in section \ref{sec:doubling}.} 
the discretisations with the SLAC derivative showed a nearly perfect behavior. 
Already at finite lattice spacing the results were very close to the correct continuum value.

Since the simulation of the full supersymmetric model demands a higher numerical effort, a sightly lower statistic was used than for the improved and unimproved model in \cite{Bergner:2007pu}.
Nevertheless $10^5$ to $4\times 10^5$ independent configurations were obtained.
The masses coincide with the improved model within the statistical errors as shown in figure \ref{fig:Mass}.
As already observed in the earlier calculations the bosonic masses show a larger statistical error than the fermionic ones.
Even the fermionic masses, that are obtained with a high precision, agree with the linear extrapolation of the masses of the improved SLAC model towards the continuum limit.
Thus the correct continuum value is obtained with the nonlocal full supersymmetric lattice realization.
In \cite{Bergner:2007pu} it was found that realizations with the SLAC derivative show an almost perfect behaviour. Only a small deviation form the continuum value is observed at finite lattice spacing.
This is also true for the model with the nonlocal interaction.

As the masses show not much difference between the considered models and do not provide a clear sign of supersymmetry breaking at a finite lattice spacing, the Ward-identities are considered.
At lower couplings and a smaller lattice spacing the Ward identities vanish in all cases.
Therefore a smaller lattice and a large coupling was chosen.
The configurations were obtained in eight independent runs with each $5\times 10^5$ independent configurations.
To estimate possible systematic errors the Ward identities of the free theories were measured, which are analytically zero.
The result is shown in figure \ref{fig:FT}, and no larger deviation from zero than $0.0003$ is observed.
This implies that if a Ward identity stays within these bounds (illustrated by a shaded region in the plot), it can be assumed to be zero within the errorbounds. 
This is not the case in the unimproved model. If the coupling constant $g$ is large enough a clear deviation from zero can be observed (figure \ref{fig:SLACU}). 
In the improved model one of the two Ward identities is zero within the errors (figure \ref{fig:SLACI}).
On the other hand the second Ward identity gets an enhanced deviation from zero.

This indicates one of the problems of a partial realized supersymmetry on the lattice: In order to avoid the breaking of one of the supersymmetries a larger breaking of the remaining supersymmetries is generically induced.
Note furthermore that the measured Ward identities of the unimproved model are related via $\mathcal{R}^{(2)}_{n}=-\mathcal{R}^{(1)}_{-n}$.
For the fermionic part of the Ward identity this is due to $\erw{\psib_x\psi_y}=-\erw{\psi_y\psib_x}$.
The bosonic part of the relation between the two Ward identities comes form the invariance of the path integral measure under the transformation $t\rightarrow -t$ in the action. 
This symmetry is present in the continuum and the  unimproved models.
Thus a partial realization of supersymmetry can only be achieved at the cost of a breaking of another symmetry of the continuum model.
Additional problems that can be induced by a partial realization of supersymmetry are discussed in \cite{Kastner:2008zc}.

A comparison of the different results obtained in \cite{Bergner:2007pu} shows that for unimproved and improved models with a symmetric derivative and a Wilson mass in the bosonic and fermionic sector the same sign of supersymmetry breaking appears.

In the full supersymmetric model both Ward identities are zero within the errors (figure \ref{fig:FS}).
These results verify that only in the full supersymmetric realization with a nonlocal interaction term supersymmetry can be realized at a finite lattice spacing.
It provides the first precise measurement of a fully supersymmetric model on the lattice. 

\section{Conclusions and outlook}
\label{sec:Conclu}
In this work I have shown that a realization of a complete supersymmetry on the lattice can only be achieved with nonlocal interactions and a nonlocal derivative.
This fact is due to the breaking of the Leibniz rule by all lattice derivative operators.
The nonlocal interaction term comes form a modified product on the lattice that restricts the momentum modes.
A concrete implementation for the realization of the nonlocal product was presented.
Another important breaking mechanism due to the fermion doubling problem is also not present in this nonlocal discretisation.
A local continuum limit without nonlocal or noncovariant counterterms can be verified in lattice perturbation theory at least in lower dimensions.
Despite the nonlocal form of the action simulations in supersymmetric quantum mechanics were performed in this work.
The realization of a complete supersymmetry on the lattice can be verified with a high precision.
In this way the present work has proven that simulations with intact supersymmetry on the lattice are indeed possible.

The general discussion of lattice perturbation theory in this work reveals a hierarchy of supersymmetry breaking mechanisms.
Each of the breaking mechanisms becomes relevant when supersymmetry breaking counterterms are needed for a supersymmetric continuum limit.
The first problem is due to the Wilson mass, that must be included in the fermionic sector for all local lattice derivative operators.
The mass parameter is, however, not present in the bosonic sector. 
This problem is relevant already in one dimension.
In the two dimensional case the differences between the fermionic and bosonic derivative operators (propagators) can also become a relevant source of supersymmetry breaking.
Both of these effects can be avoided when the same derivative and mass operators are chosen for fermions and bosons.
In the on-shell theory this requires also an unconventional modification of some bosonic vertices.
In lattice perturbation theory the missing Leibniz rule becomes a relevant effect in the four-dimensional models.
This makes a full supersymmetric realization attractive especially in this case, where, however, it still has to be proven that no nonlocal counterterms are needed in the continuum limit.

On the nonperturbative level it is still not guaranteed that the violation of supersymmetry can be neglected even in lower dimensions.
Thus a model with complete supersymmetry on the lattice is a good cross check for other simulations.
If the simulations of a nonlocal model with intact lattice supersymmetry and a local model without lattice supersymmetry leads to the same results, the recovery of locality and supersymmetry in the continuum limit can be ensured.

On the technical side the simulations of the nonlocal lattice action require a large numerical effort. 
The formulation used in this work provides a possibility only in the lower dimensional case.
Further improvements are, however, possible. 
The nonperiodic delta function \eq{eq:nonperdelta} is realized via a periodic delta function on a finer lattice.
One can also formulate the whole lattice theory on the finer lattice, where no modification of the product is necessary.
On the finer lattice the only important requirement for the full supersymmetric theory is a restriction of all momentum modes below the cutoff of the coarser lattice.
In addition one has to keep in mind that the theory effectively lives on a coarser lattice when the observables are computed.
This opens another possibility for the calculation of the supersymmetric theory.
The lower cutoff of the theory may be implemented via a momentum-truncation of the force, such that no higher momenta are generated in the update step.
This must be ensured for (pseudo-) fermionic and bosonic fields.
Instead of a sharp cutoff one can also try to realize a smooth version of it.
Instead of the full supersymmetric model one can also try to find improved actions with a reduced violation of the Leibniz rule and a moderate suppression of higher momentum modes.\footnote{A similar treatment was suggested in \cite{Fujikawa:2002ic}.}

A disadvantage of the present approach is that it can not be directly applied for supersymmetric gauge theories.
The SLAC derivative requires a gauge invariant connection between distant lattice points and according to \cite{Karsten1979100} nonlocal and noncovariant counterterms are needed in the continuum limit.
Perhaps a gauge fixed version of this theory can be used in the simulations.
\section*{Acknowledgments}
I thank A.~Wipf for corrections and helpful discussions; T.~K\"astner and C.~Wozar for the assistance with the development of the simulation code.
I also like to thank F.~Bruckmann and U.~Theis for interesting discussions.
I acknowledge support by the Evangelisches Studienwerk. 
\begin{appendix}
\section{Superspace formulation and off-shell representation of the one dimensional theory}
\label{app:superspace}
The one-dimensional (bosonic) superfield has the following expansion in the Gra\3mann coordinates $\t$ $\tb$
\begin{equation}
 \P(t,\t,\tb)=\vp +\tb\psi+\psibar\t +\tb \t F\, .
\end{equation}
It contains the real bosonic field $\vp$, the fermions $\psi$ and $\psibar$ as well as the auxiliary field $F$.
The supersymmetry transformations of the superfields are generated by the supercharges
\begin{equation}
 Q=i\pa_{\tb}+\t \pa_t\,\quad \text{and} \quad \Qb=i\pa_{\t}+\tb \pa_t\, .
\end{equation}
This implies the following supersymmetry transformation for the component fields:
\begin{align}
 \d\vp&=i\vepsb\psi-i\psibar\veps\,,\quad \d\psi=(\pa_t\vp-iF)\veps\,,\quad\nonumber\\
 \d\psibar&=\vepsb(\pa_t\vp+iF)\,, \quad \d F=-\vepsb\pa_t\psi-\pa_t\psibar\veps\, .
\end{align}
A Lagrangian is constructed using the covariant derivatives 
\begin{equation}
 D=i\pa_{\tb}-\t \pa_t\,, \quad \Db=i\pa_{\t}-\tb\pa_t\, ,
\end{equation}
that anticommute with $Q$ and $\Qb$.
The action is hence obtained from an integration over the whole superspace,
\begin{eqnarray}
\label{eq:contActionSusyQM}
 S&=&\inte\dttb dt \left[ \half \P(t,\t,\tb) K \P(t,\t,\tb)+iW(\P(t,\t,\tb))\right]\\
&=&\inte dt \left( \half (\pa_t\vp)^2-i\psib\pa_t\psi+\half F^2+i FW'(\vp)-i\psib W''(\vp)\psi\right)\, ,
\end{eqnarray}
where $W(\P)$ (here $W(\P)=m\P^2/2+g\P^4/4$)  is a polynomial in $\P$ and $K=\half (D\Db -\Db D)$.
\section{Some conventions}
\label{sec:Conv}
In the $\mu$ direction there are assumed to be $N_\mu$ lattice points separated by the spacing $a_\mu$.
The number of lattice points is odd in each direction.
If the index of $N$ and $a$ is not specified the same number of points and the same lattice spacing applies for all directions. 
Each lattice point is labelled by a vector $n$ with $\D$ integer components running from $0$ to $N_\mu-1$.
The lattice point is defined as $x_n=\sum_\mu (n)_\mu a_\mu$.
$\vp_n$ is the value of the field at this point.
The vector $e_\mu$ has zero components except a $1$ in its $\mu$ direction ($x_{n+e_\mu}$ is the next neighboring lattic point of $x_n$ in $\mu$ direction).

The size of the lattice in $\mu$ direction is, consequently, $L_\mu=N_\mu a_\mu$ and its volume $\Omega_L=\prod_\mu L_\mu$.
Periodic boundary conditions are assumed for this volume.  
If not further specified the sum $\sum_n$ of a lattice index is
\begin{equation}
 \sum_n=\prod_\mu\left( a_\mu \sum_{(n)_\mu=0}^{(n)_\mu=N_\mu-1}\right)
\end{equation}

The lattice index is labeled by $n$ and $m$. $n_i$ or $n_j$ are treated as individual lattice indices; only the $(n)_\mu$ or $(n)_\nu$ stands for a component of an index. 
Thus $(n_1)_\mu$ is the component of $n_1$ in $\mu$ direction. The same applies for the dimensionless wave vector $k$.

The derivative operators used here are defined as
\begin{equation}
 \begin{array}{lc}
  \text{symmetric derivative}&(\symder_\mu \vp)_n=\frac{1}{2a_\mu}(\vp_{n+e_\mu}-\vp_{n-e_\mu}) \\
\text{forward derivative}&(\leftder_\mu\vp)_n=\frac{1}{a_\mu}(\vp_{n+e_\mu}-\vp_{n})\\
  \text{backward derivative}&(\rightder_\mu \vp)_n=\frac{1}{a_\mu}(\vp_{n}-\vp_{n-e_\mu})\\
\text{SLAC derivative}&(\derSLAC_\mu\vp)_n=\frac{\pi}{a_\mu N_\mu}\sum_{l=0}^{N_\mu-1} (-1)^{(n)_\mu-l}\frac{\vp_{n+le_\mu }}{\sin(\pi((n)_\mu-l)/N_\mu)}
 \end{array}
\label{eq:derOps}
\end{equation}
The SLAC derivative\footnote{This derivative is also called DWY derivative (Drell, Weinstein, and Yankielowicz).}, \cite{Drell:1976bq,Drell:1976mj,Kirchberg:2004vm}, is derived from a discretisation of the Fourier space representation of the continuum derivative operator. 

Functions on the lattice can be represented in Fourier space according to,
\begin{eqnarray}
 \phi_n&=&\sum_k\phi(p_k)e^{i p_k x_n}\\
 \phi(p_k)&=&\sum_{n}\phi_n e^{-i p_k x_n}\, ,
\end{eqnarray}
with $(p_k)_\mu=\frac{2 \pi k_\mu}{L_\mu}$  and the components of $k$ are integers running form $-(N_\mu-1)/2$ to $(N_\mu-1)/2$. ($p_q x$ here stands for a scalar product of the two vectors.)
If not further specified the above sum over $k$ represents
\begin{equation}
 \sum_k=\frac{1}{\Omega_L}\prod_\mu\sum_{(k)_\mu=-(N-1)/2}^{(k)_\mu=(N-1)/2}\, .
\end{equation}
$\phi(p_k)$ is periodic in $p_k$, $\phi(p_k)=\phi(p_k+e_\mu l 2\pi/a_\mu)$ $\forall l\in
\mathbb{Z}$ and all directions $\mu$. 
The momentum $(p_k)_\mu$ of the modes is inside the Brillouin zone (BZ)  defined by $\mbox{BZ}=\{(p_\mu)|\,|p_\mu|\leq(\Lambda_L)_\mu=\frac{\pi}{a_\mu}\}$.

The Fourier transformation implies the following representation of the delta on the lattice
\begin{eqnarray}
 \d_L(x_n-x_m)&:=&\sum_k e^{ip_k(x_n-x_m)}=\prod_\mu (a_\mu)^{-1} {\bar\d}_{(n)_\mu m_\mu}\nonumber\\
 \d_L(p_{k_1}-p_{k_2})&:=&\sum_n e^{-i(p_{k_1}-p_{k_2}) x_n}=\prod_\mu (N_\mu a_\mu) {\bar\d}_{(k_1)_\mu (k_2)_\mu}\, ,
\label{eq:periodicDelta}
\end{eqnarray}
where $\bar\d_{(n)_\mu m_\mu}$ is one for $(n)_\mu=m_\mu\mod N_\mu$ and zero otherwise. The two delta functions are periodic: $\d_L(p_k)=\d_L(p_k+e_\mu l 2\pi/a_\mu)$; $\d_L(x_k)=\d_L(x_k+e_\mu l N_\mu)$ $\forall l\in
\mathbb{Z}$ and $\mu$.

For the lattice perturbation theory the thermodynamic of the lattice expressions is performed.
Then one gets the following Fourier representation with the now continuous  momentum $p$:
\begin{eqnarray}
 \phi_n&=&\int_p  \phi(p)e^{i p x_n}:=\int_\text{BZ} \frac{d^D p}{(2\pi)^D}\; \phi(p)e^{i p x_n}\\
 \phi(p)&=&\sum_{n}\phi_n e^{-i p x_n}\, .
\end{eqnarray}
($n \in \mathbb{Z}^\D$ now runs over an infinite number of lattice points.)
One can easily change between the Fourier representation on a finite lattice and in the thermodynamic limit.
The expressions in Fourierspace remain the same, only the discrete momentum $p_k$ has to be replaced by the continuous $p$. 
This continuous momentum is, because of the finite lattice spacing still restricted to the BZ.
Consequently the delta in Fourier space is still periodic as on the lattice.

In a similar way the Fourier representation of translational invariant operators with two indices is derived on the lattice:
\begin{eqnarray}
  \n^\mu_{nm}&=&\sum_k\n^\mu(p_k)\,
  e^{i p_k (x_n-x_m)}\, .
\end{eqnarray}
The obtained operator has the same periodicity as the Fourier space representation of a field.
On the lattice the matrix entries of $\nabla$ should be real.
Therefore, the imaginary part of $\nabla(p)$ is antisymmetric ($\Im\nabla(-p)=-\Im\nabla(p)$) and the real part is symmetric ($\Re\nabla(-p)=\Re\nabla(p)$).

An antisymmetric derivative operator acting in $\mu$ direction can always be represented as
\begin{equation}
  \n_{nm}^\mu=\sum_{r=1}^{N-1}c_r (\n^{(r)})^\mu_{mn}\, , \quad \text{with} \quad (\n^{(r)})^\mu_{nm}=\d_{n+re_\mu,m}-\d_{n-re_\mu,m}\, ,
\label{eq:asymrep}
\end{equation}
and some constants $c_r$.

For the operators defined in \eq{eq:derOps} the Fourier representation is thus
\begin{eqnarray}
 \symder_\mu(p)&=&\frac{i}{a_\mu}\sin(p_\mu a_\mu)\nonumber\\
 (\leftder_\mu\rightder_\mu)(p)&=&\frac{4}{a_\mu^2}\sin^2(p_\mu a_\mu/2)\nonumber\\
 \MW(p)&=&\sum_\mu\frac{2r}{a_\mu}\sin^2(p_\mu a_\mu/2)\nonumber\\
\derSLAC_\mu(p)&=&ip_\mu\, .
\end{eqnarray}
Apart from the SLAC derivative these are all functions of the type $a^d F(a p)$ where $d$ is determined by the dimension of the operator.
Obviously the behavior of $F$ in the vicinity of $p=0$ is the important part in the continuum limit unless there are other points with $F=0$.

Now consider a more complicated operator $\Ct_{m_1,\ldots m_{\nf}}$. Translational invariance implies that a shift of  all lattice points by $x_m$ is irrelevant.
One way for the representation in Fourier space is
\begin{eqnarray}
 \Ct_{m_1,\ldots,m_{\nf}}&=&\sum_{k_1,\ldots,k_{\nf}}\Ct(p_{k_1},\ldots,p_{k_{\nf}})e^{i (p_{k_1}x_{m_1}+\ldots+p_{k_{\nf}}x_{m_{\nf}})}\nonumber\\
 \Ct(p_{k_1},\ldots,p_{k_{\nf}})&=&\sum_{m_1,\ldots,m_{\nf}}\Ct_{m_1,\ldots,m_{\nf}}e^{-i (p_{k_1}x_{m_1}+\ldots+p_{k_{\nf}}x_{m_{\nf}})}\, .
\label{eq:repFSCt}
\end{eqnarray}
The translational invariance means for the Fourier space representation
\begin{equation}
 \Ct(p_{k_1},\ldots,p_{k_{\nf}})e^{i(p_{k_1}+\ldots+p_{k_{\nf}})x_m}=\Ct(p_{k_1},\ldots,p_{k_{\nf}})\, .
\end{equation}
This implies $p_{k_{\nf}}^\mu=-p_{k_1}^\mu-\ldots-p_{k_{\nf-1}}^\mu \mod \Lambda_L^\mu$ for all $\mu$.
A representation that explicitly implies the translational invariance is obtained by going from $\Ct$ with $\nf$ indices to a matrix $C$ with $\nf-1$ indices according to
\begin{equation}
 \Ct_{m_1,\ldots,m_{\nf}}=C_{(m_{\nf}-m_1),\ldots,(m_{\nf}-m_{\nf-1})} \, .
\end{equation}
A representation in Fourier space of this matrix is
\begin{align}
 C_{m_1,\ldots,m_{{\nf}-1}}=\!\!\!\!\!\!\sum_{k_1,\ldots,k_{{\nf}-1}}\!\!\!\!\!C(p_{k_1},\ldots,p_{k_{{\nf}-1}})e^{i p_{k_1}x_{m_1}+\ldots+i p_{k_{{\nf}-1}}x_{m_{{\nf}-1}}}.
\label{eq:repFSC}
\end{align}
Thus the two representations are in Fourierspace related by
\begin{equation}
 \Ct(p_{k_1},\ldots,p_{k_{\nf}})=\d_L(p_{k_1}+\ldots+p_{k_{\nf}})C(p_{k_1},\ldots,p_{k_{{\nf}-1}})\, .
\end{equation}
With these operators the modification of the product on the lattice \eq{eq:ModProd} is defined. 

The commutativity of the product is ensured by the fact that $\Ct$ is invariant under the exchange of its arguments.
A requirement for a modified product, not discussed in the main text, is its associativity,
\begin{equation}
 (\one{\p}\ast\two{\p}\ast\p^{(3)})_l=((\one{\p}\ast\two{\p})\ast\p^{(3)})_l.
\end{equation}
In Fourier space associativity on the lattice demands
\begin{align}
 \Ct(p_{k_1},p_{k_2},p_{k_3},p_{k_4})&=\sum_{k_5} \Ct(p_{k_5},p_{k_2},p_{k_3})\Ct(p_{k_1},-p_{k_5},p_{k_4})\\
 \text{or}\quad C(p_{k_1},p_{k_2},p_{k_3})&=\sum_{k_4}\d_L(p_{k_4}-p_{k_2}-p_{k_1}) C(p_{k_4},p_{k_3})C(p_{k_1},p_{k_2}).
\end{align}
This condition is fulfilled for the proposed lattice action in section \ref{sec:Real}.

The matrix $\Fc$ used in this construction follows from
\begin{eqnarray}
\Fc_{nm}&=& \frac{1}{aN}\sum_{k=-(N-1)/2}^{(N-1)/2} \exp\left(i\frac{2\pi k}{aN}\left(am-\frac{a}{(n_f-1)}n\right) \right)\nonumber\\
&=&\frac{e^{-i\frac{2\pi(N-1)}{2N}(m-\frac{n}{n_f-1})}}{aN}\frac{1-e^{i2 \pi(m-n/(n_f-1))}}{1-e^{i2 \pi/N(m-n/(n_f-1))}}\nonumber\\
&=&\frac{\sin(\pi(m-n/(n_f-1))}{aN\sin(\pi/N(m-n/(n_f-1))}\, .
\label{eq:FCDef}
\end{eqnarray}
It is clear that for this kind of matrix the following summation rule holds
\begin{equation}
\label{eq:sumFc}
 \frac{a}{(n_f-1)}\sum_{n=0}^{n_f N-1}\Fc_{nm_1}\Fc_{nm_2}=\d(x_{m_1}-x_{m_2})\, .
\end{equation}
This implies for the modified product of two fields
\begin{equation}
 \sum_l (\one{\p}\ast\two{\p})_l=\sum_l \one{\p}_l \two{\p}_l\, .
\end{equation}

$\Fc$ maps the fields $\vp^{(i)}$ with a lattice size $N$ on the fields $\tilde{\vp}^{(i)}_n=\sum_n\Fc_{nm}\vp^{(i)}_n$ with a lattice size $(n_f-1)N$.
The fields on the larger lattice have, however, a momentum constraint that is constraint below $\frac{\pi}{a}$ instead the larger lattice cutoff $\frac{\pi(n_f-1)}{a}$.
Thus $\Fc$ generates a one to one map of $\vp^{(i)}(p_k)$ onto the modes of the larger lattice below the cutoff $\frac{\pi}{a}$.
\end{appendix}

\begingroup\raggedright

\endgroup
\end{document}